\documentclass{article}
\usepackage{fancyhdr}

\usepackage{report}

\usepackage{silence}
\usepackage{soul}
\usepackage[utf8]{inputenc}
\usepackage[T1]{fontenc}
\usepackage{hyperref}
\usepackage{xurl}
\usepackage{booktabs}
\usepackage{amsfonts}
\usepackage{fancyhdr}

\usepackage{nicefrac}
\usepackage{microtype}
\usepackage{graphicx}
\usepackage{fontawesome5}
\usepackage{pifont}
\usepackage{multirow}
\usepackage{makecell}
\usepackage{CJKutf8}
\definecolor{darkmagenta}{rgb}{0.56, 0.0, 1.0}
\definecolor{softyellow}{rgb}{1.0, 0.92, 0.3}
\definecolor{LightAquamarine}{rgb}{0.75, 1.0, 0.8}
\definecolor{FireBrick}{RGB}{178,34,34}
\definecolor{MediumPurple}{RGB}{147,112,219}
\usepackage{xcolor}

\definecolor{uclablue}{rgb}{0.15, 0.45, 0.68}
\hypersetup{
    breaklinks,
    colorlinks=true,
    citecolor={darkmagenta},
    linkcolor={uclablue},
    urlcolor={uclablue}
}
\usepackage{wrapfig}
\usepackage{float}
\usepackage{placeins}
\usepackage{subcaption}
\usepackage{needspace}
\usepackage{tcolorbox}
\usepackage{amsmath}
\usepackage{amssymb}
\usepackage{xspace}
\usepackage{enumitem}
\tcbuselibrary{breakable}
\usepackage{colortbl}
\usepackage{transparent}

\newcommand{\sidewrapwidth}{0.48\textwidth}

\definecolor{headerblue}{RGB}{40, 80, 140}
\definecolor{rowgray}{RGB}{248, 248, 252}
\definecolor{highlight}{RGB}{230, 210, 255}
\definecolor{best}{RGB}{0, 140, 100}
\definecolor{second}{RGB}{30, 100, 180}
\definecolor{partialorange}{RGB}{200, 120, 20}
\newcommand{\best}[1]{\textcolor{best}{\textbf{#1}}}
\newcommand{\second}[1]{\textcolor{second}{\underline{#1}}}

\newcommand{\mfull}{\textcolor{best}{\ding{51}}}
\newcommand{\mpart}{\textcolor{partialorange}{$\bigcirc$}}
\newcommand{\mno}{\textcolor{red!70}{\ding{55}}}
\newcommand{\mna}{\textcolor{black!45}{---}}

\definecolor{tbdred}{RGB}{200, 30, 40}
\newif\ifshowtbd
\showtbdtrue

\definecolor{promptbg}{RGB}{248,249,252}
\definecolor{promptframe}{RGB}{52,110,183}
\tcbuselibrary{most,listings}
\newtcblisting{promptbox}[1]{%
  listing only, colback=promptbg, colframe=promptframe,
  title={\small\bfseries #1}, fonttitle=\small\bfseries,
  listing options={basicstyle=\ttfamily\scriptsize, breaklines=true,
    breakautoindent=false, breakindent=0pt, breakatwhitespace=true,
    postbreak=\mbox{\textcolor{gray}{$\hookrightarrow$}\space},
    columns=fullflexible, keepspaces=true,
    extendedchars=true, inputencoding=utf8,
    literate={—}{{---}}3 {–}{{--}}2 {‘}{{`}}1 {’}{{'}}1 {“}{{``}}2 {”}{{''}}2 {…}{{...}}3,
    aboveskip=0pt, belowskip=0pt},
  boxrule=0.6pt, arc=2pt, left=6pt, right=6pt, top=2pt, bottom=2pt, breakable,
}

\definecolor{njuPurple}{RGB}{220,205,230}
\definecolor{njuPurpleLight}{RGB}{250,245,252}
\newtcolorbox{abstractbox}{
    colback=njuPurpleLight, colframe=njuPurple, boxrule=1pt, arc=4mm,
    left=8pt, right=8pt, top=8pt, bottom=8pt, opacityback=0.95
}

\newcommand{\benchname}{GameLogicBench\xspace}

\title{\benchname: Evaluating Coding Agents on Runtime\\ Game Logic with Tick-Level State Assertions}

\author{
  Xinyu Che$^{*}$ \quad Yunfei Ge$^{*}$ \quad Shihao Li$^{*}$  \quad Yanchen Liu$^{*}$ \quad Hang Yan$^{*}$ \quad Xinping Lei$^{*}$ \\
  Yanghai Wang \quad Zixuan Dong \quad Yifan Yao \quad Qianqian Xie \quad Letian Zhu \quad Jiaheng Liu$^{\dagger}$
  \\
\vspace{4mm}
\large
\textbf{Nanjing University}
\\
\vspace{2mm}
\texttt{kosmoche@gmail.com}, \texttt{liujiaheng@nju.edu.cn}
}

\begin{document}

\maketitle
\begingroup
\renewcommand{\thefootnote}{}
\footnotemark
\footnotetext{*~Equal Contribution. ~~$^\dagger$~Corresponding Author.\\
  \hspace*{1.8em}\faGithub~\url{https://github.com/NJU-LINK/GameLogicBench}}
\addtocounter{footnote}{-1}
\endgroup

\vspace{-8mm}
\begin{abstractbox}
\begin{center}
\textbf{\Large Abstract}
\end{center}
Coding agents can now modify and test code across large software projects. Game development is one such domain, where agents must implement gameplay rules. A game can end in a valid state even after violating its rules during the run. Current game-development benchmarks replay fixed examples, score videos, or ask another model to judge the result. However, no existing benchmark checks game rules throughout execution across varied evaluator-selected scenarios while ensuring exactly reproducible verdicts.
We introduce \benchname, a benchmark of 72 gameplay-logic tasks in Godot projects. An automated evaluator checks each game's rules at every simulation tick. Across 403 hand-designed scenarios, seeded parameter variations produce 1,451 test cases.
To ensure that the evaluator measures behavior rather than implementation choice, it must accept different correct implementations for each task while rejecting mutants, implementations with one required capability removed.
The tasks span isolated mechanics, multi-system interactions, and repository-scale features. Across 20 combinations of language models and scaffolds, the best observed run solves 52.78\% of tasks. Under Claude Code, all twelve models solve fewer tasks as task scope expands from isolated mechanics, through interacting systems, to repository-scale features.
Agents inspect code more often and make more tool calls on repository-scale tasks than on isolated-mechanic tasks. Most unsuccessful submissions are runnable, but implement some required game behavior incorrectly. We compared versions of our benchmark evaluator built with and without validation using mutants. Without this validation, some incorrect agent submissions passed. A separate analysis finds agents copying code from the original public repositories when network access is open. Reliable evaluation thus depends both on what the tests reject and on what external code agents can access.
\end{abstractbox}

\begin{figure}[H]
\centering
\includegraphics[width=\textwidth]{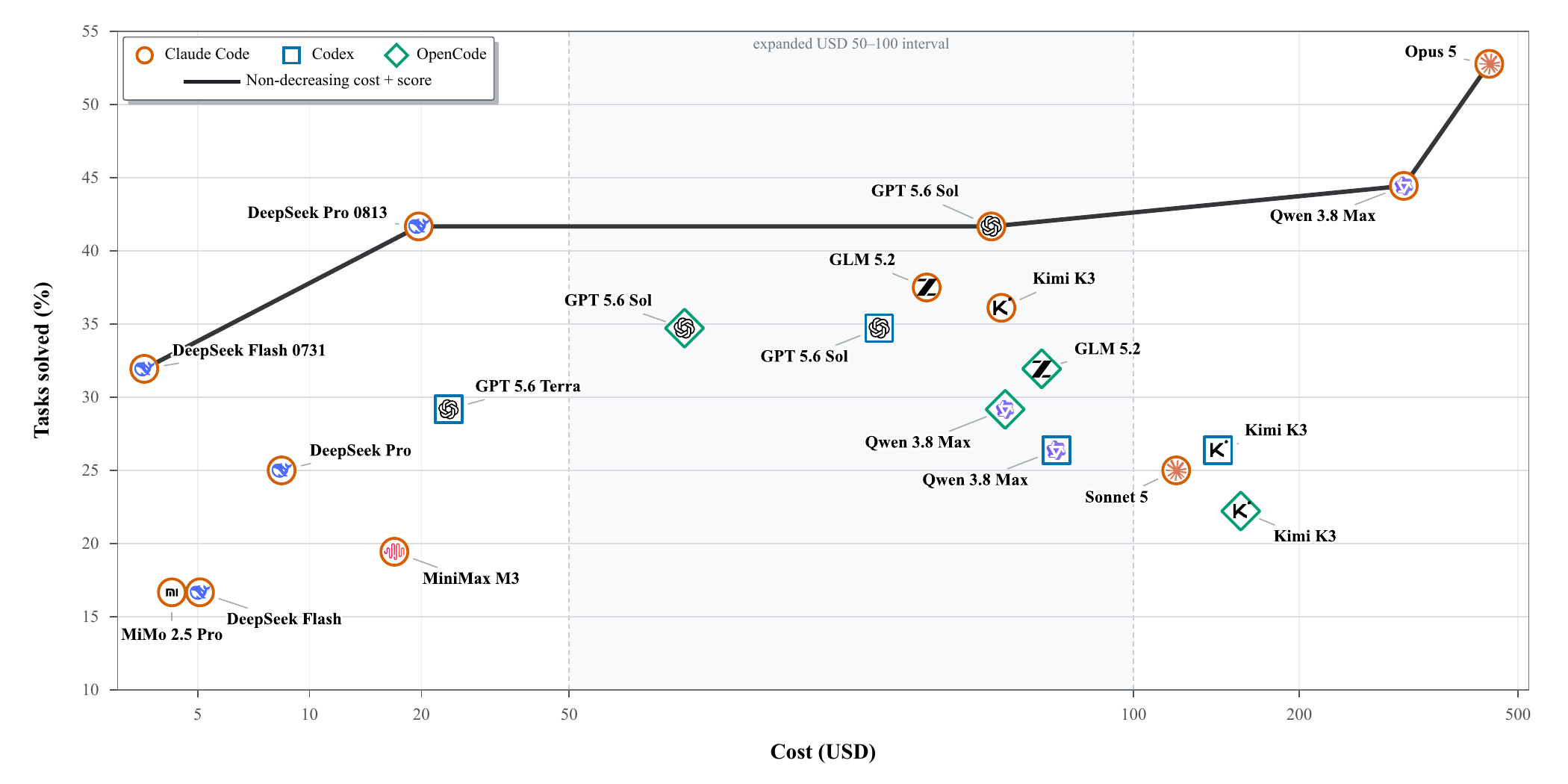}
\caption{Observed solve rate versus total inference cost by configuration.}
\label{fig:main_results_cost}
\end{figure}
\section{Introduction}
\label{sec:intro}

Coding agent benchmarks have expanded from isolated function synthesis to repository-level issue resolution in SWE-bench~\citep{swebench2024} and challenging terminal tasks in Terminal-Bench~\citep{terminalbench2026}. These benchmarks assess task completion using automated tests. Game development has recently emerged as a new domain for evaluating coding agents~\citep{gamedevbench2026,gameenginebench2026,webgamebench2026,gamecraft2026}. Agents must implement gameplay rules within interacting systems. Game development thus presents the unique challenge of maintaining correct behavior across every simulation tick.

\begin{figure}[ht]
\begin{center}
\includegraphics[width=\linewidth]{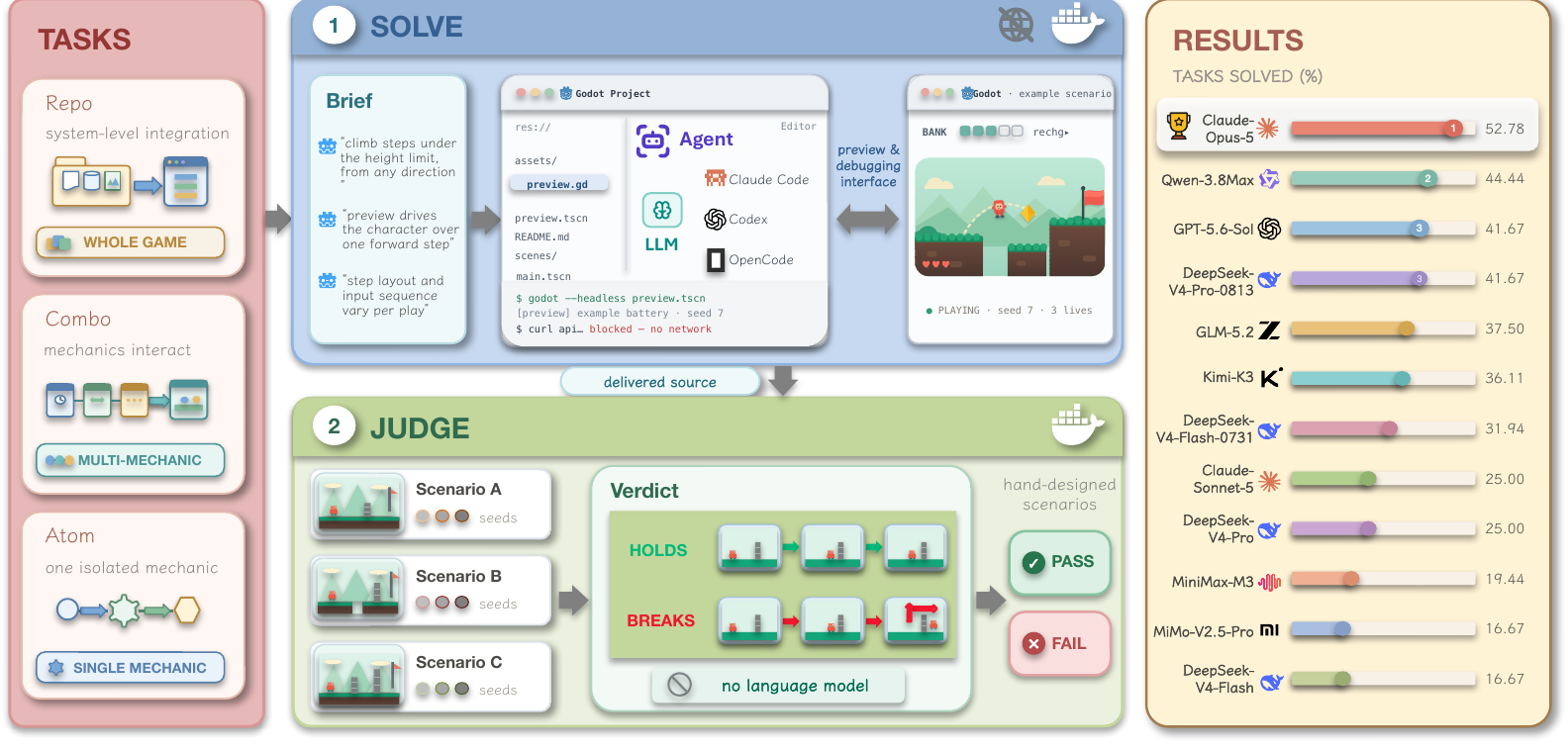}
\end{center}
\caption{\benchname{} at a glance. \textbf{Left:} task tiers by integration scope. \textbf{SOLVE:} agents work in a sealed Godot container. \textbf{JUDGE:} a deterministic judge evaluates solutions across hand-designed scenarios and seeds without a language model. \textbf{Right:} solve rates under Claude Code.}
\label{fig:overview}
\end{figure}

\begin{table*}[htbp]
\caption{Benchmarks whose criteria judge a running artifact, grouped by substrate. \mfull~indicates all tasks, \mpart~some tasks, \mno~no tasks, and \mna~not applicable. \textbf{Determ.} denotes whether the discriminating score component is free of model judgment; \textbf{Tick-level} whether assertions read game state at two or more simulation ticks; and \textbf{Scenarios} gives the number of hand-designed scenarios per task.}
\label{tab:benchmark_comparison}
\begin{center}
\resizebox{\textwidth}{!}{%
\renewcommand{\arraystretch}{1.3}
\begin{tabular}{@{}lllllcccc@{}}
\toprule
\textbf{Benchmark} & \textbf{Runtime} & \textbf{Task} & \textbf{Codebase}
& \textbf{Evaluation} & \textbf{Determ.} & \textbf{Tick-level} & \textbf{Scenarios} & \textbf{Size} \\
\midrule
\rowcolor{rowgray}
\multicolumn{9}{@{}l}{\textit{\textcolor{headerblue}{Software repositories}}} \\
SWE-bench~\citep{swebench2024} & \mna & Patch & Real project
& Unit tests & \mfull & \mna & \mna & 2{,}294 \\
Terminal-Bench 2.0~\citep{terminalbench2026} & \mna & Both & Both
& Unit tests & \mfull & \mna & \mna & 89 \\
\midrule
\rowcolor{rowgray}
\multicolumn{9}{@{}l}{\textit{\textcolor{headerblue}{Interactive web artifacts}}} \\
WebCompass~\citep{webcompass2026} & Chromium / JS & Both & Both
& Agent judge & \mno & \mna & 1 & 1{,}526 \\
WebGameBench~\citep{webgamebench2026} & Chromium / JS & From-scratch & Purpose-built
& Agent judge & \mno & \mno & 1 & 111 \\
\midrule
\rowcolor{rowgray}
\multicolumn{9}{@{}l}{\textit{\textcolor{headerblue}{Real game engines}}} \\
GameDevBench~\citep{gamedevbench2026}       & Godot / GDScript & Patch & Real project
& Test scripts & \mfull & \mpart & 1 & 333 \\
JAMER~\citep{jamer2026}                     & Godot / GDScript & Both & Both
& Engine verify & \mfull & \mno & 1 & 300 \\
GameCraft-Bench~\citep{gamecraft2026}       & Godot / GDScript & From-scratch & Purpose-built
& VLM judge & \mno & \mno & 1 & 140 \\
GameEngineBench~\citep{gameenginebench2026} & UE5 / C++ & Patch & Real project
& Tests + LLM judge & \mno & \mpart & 1 & 110 \\
AutoUE~\citep{autoue2026}                   & UE5 / C++ & From-scratch & Purpose-built
& LLM judge & \mno & \mno & 1 & 20 \\
\midrule
\rowcolor{highlight}
\textbf{\benchname (ours)} & Godot / GDScript & Patch & Both
& Engine assertions & \mfull & \mfull & 2--12 & 72 \\
\bottomrule
\end{tabular}%
}
\end{center}
\end{table*}

However, existing game-development benchmarks do not consistently check whether game rules hold throughout execution. A run may finish in a seemingly valid state despite violating game rules during execution. Table~\ref{tab:benchmark_comparison} compares existing coding agent benchmarks and their evaluation methods. With one fixed evaluation scenario, a solution may pass the test yet fail under other valid inputs or layouts~\citep{gamedevbench2026,gameenginebench2026}. GameCraft-Bench evaluates gameplay recordings with a multimodal judge~\citep{gamecraft2026}. The evaluated agent records and selects these videos, so failures outside the submitted recordings may go unobserved. Video-based judgments can also miss rule violations that are not visible in the rendered frames. Other benchmarks use agent judges to interact with the submitted application and grade it against the requirements~\citep{webcompass2026,webgamebench2026,autoue2026,playcoder2026}. The evaluator chooses the interactions to test and assigns the score using model judgments, so repeated evaluations may produce different verdicts. Such evaluations are also expensive. The judge that agrees best with human raters in WebCompass costs \$4.66 per sample~\citep{webcompass2026}, and WebGameBench budgets up to two hours of evaluator rollout per game~\citep{webgamebench2026}. None of the compared benchmarks combines \emph{multiple judge-selected scenarios}, \emph{tick-level assertions across all tasks}, and \emph{deterministic verdicts}.

To address the above-mentioned problems, we introduce \benchname, a benchmark of 72 gameplay-logic tasks. Each task asks the agent to implement one mechanic inside an existing Godot project. Our evaluation combines these three properties. The judge scores a solution across several hand-designed scenarios of its own choosing (\emph{multiple judge-selected scenarios}). A \emph{tick} is one discrete update of the game simulation. \emph{Tick-level assertions} check the game's internal state and event history at simulation ticks against the task's behavioral requirements. Our judge applies these checks at every tick of every task's run, whereas industrial systems typically automate gameplay testing at a coarser granularity~\citep{paredes2019division}. Scoring involves no language model, so any reader can replay a verdict exactly (\emph{deterministic verdicts}).

We source candidate mechanics from open-source Godot repositories, released games, and a taxonomy of functional capabilities. Human annotators screen them for deterministic evaluation, after which agents build the task projects and judges. Each task must pass calibration against correct implementations and mutants, independent verification, and human review before admission. Across 20 model--scaffold configurations, the best observed solve rate is 52.78\%. In 74.3\% of failed scenarios, the submitted solution runs but violates required game behavior. These results show that current coding agents still struggle to implement gameplay logic correctly. Our ablations show that final-state checks can overestimate task success by accepting solutions that violate game rules during execution.

Our contributions are as follows:
\begin{itemize}
    \item We identify limitations in existing game-development evaluations and show that
    restricted scenario coverage and final-state checks can overestimate task success.
    \item \benchname{} provides a deterministic, tick-level benchmark that
    tests whether gameplay logic remains correct throughout execution.
    \item We validate each task criterion against correct implementations and mutants. Ablations show that this validation is necessary for
    reliable scoring. We also find that most failed scenarios involve runnable
    solutions that violate game rules, and that open-network source retrieval
    can inflate repository-tier scores.
\end{itemize}

\section{\benchname}
\label{sec:benchmark}

Game mechanics are the rules that govern how characters and objects behave, such as whether a character can climb a step.
\benchname organizes gameplay-logic tasks into three tiers of integration scope.
\emph{Atom tasks} isolate one mechanic in a minimal game developed for the benchmark.
\emph{Combo tasks} combine several mechanics in a game developed for the benchmark, where correctness depends on their interactions in time, in space, and under concurrent calls.
\emph{Repo tasks} require a mechanism to be implemented in place inside a real project whose heterogeneous subsystems maintain whole-game invariants over shared state.

\subsection{Task Definition}
\label{subsec:anatomy}

Each task provides the agent with a brief and a Godot project that includes a preview and debugging interface. The brief states the behavioral and interface requirements and names the files the agent may modify. The preview runs one public example scenario. Evaluation runs the submitted solution through the same interface across multiple scenarios. The rest of the project is the game in which the required behavior must be implemented.

For example, in one repo-level task, the agent must rebuild a character-movement component while preserving the interface used by the rest of the project. The brief requires the controller to climb steps at or below a configured height from any travel direction, without treating taller obstacles as steps. In the public preview, the character approaches one step from \(+X\). In a scored scenario, it instead approaches from \(-X\), with a wall above the step limit farther along the path. The required behavior and interface stay fixed while the layout and input sequence change. The full brief for this task is provided in Appendix~\ref{app:task_briefs}.

Each task has a three-layer information boundary. First, the brief discloses the functional requirements and the game's rules, including that the scenario is rebuilt on every run. Second, it does not list the scored scenarios, which exercise the solution through the preview's interface while varying the layout and input sequence. Third, the brief omits the judge implementation, keeping the task focused on the required behavior rather than how that behavior is scored.

\begin{figure}[htbp]
\begin{center}
\includegraphics[width=\textwidth]{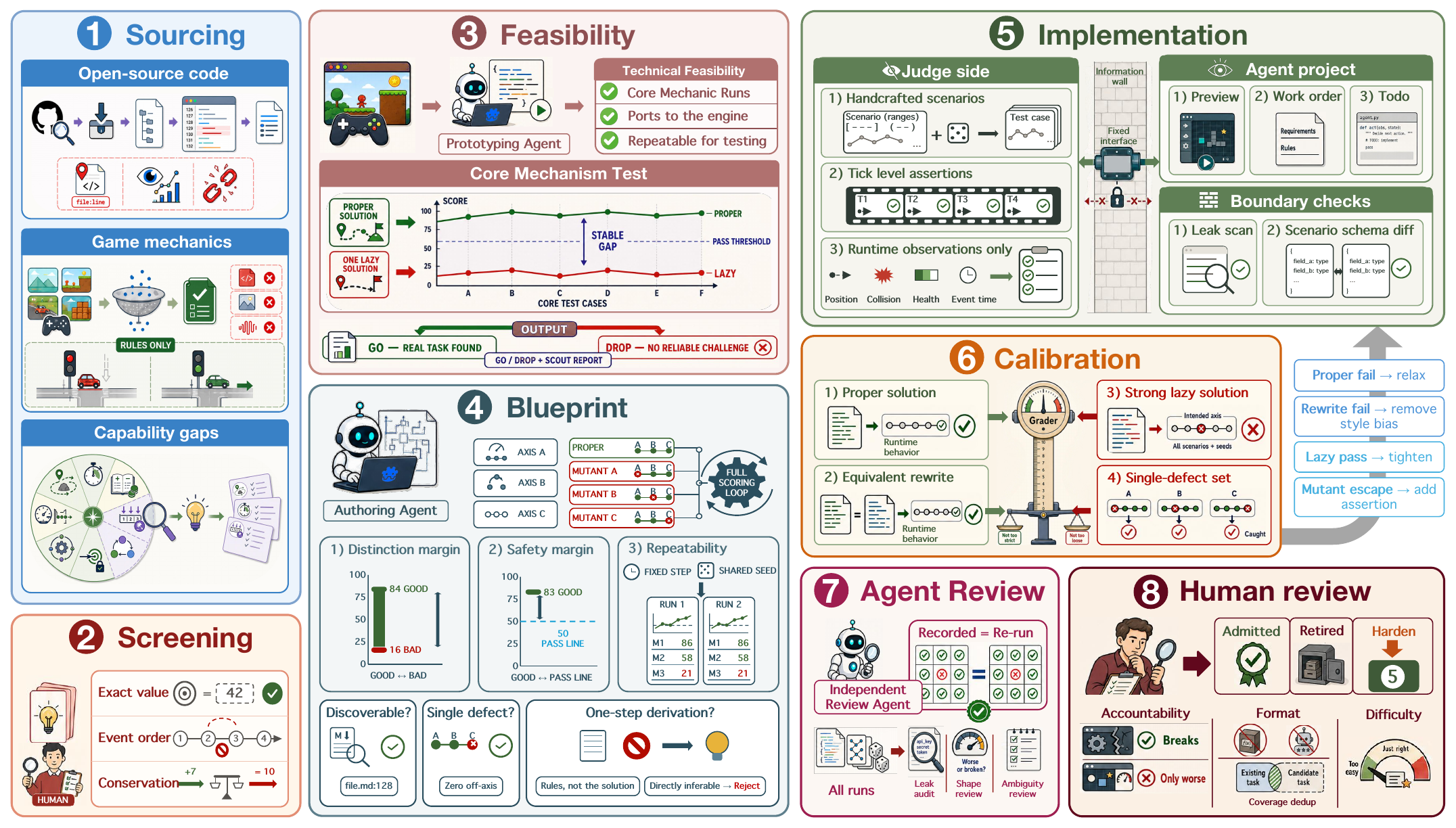}
\end{center}
\caption{Overview of the benchmark-task construction and verification pipeline.}
\label{fig:benchmark_construction}
\end{figure}

\subsection{Benchmark Construction}
\label{subsec:construction}

The pipeline is curated by humans at both ends, with agent-assisted construction and verification in between. Its stages are ordered from cheap to expensive, as shown in Figure~\ref{fig:benchmark_construction}.

\textbf{Sourcing.}
We collect candidate game mechanics from open-source Godot repositories, released games, and a taxonomy of functional capabilities. We reuse repository code when its license permits reuse. Otherwise, we implement the mechanic in a new game for the benchmark. For mechanics drawn from released games, we build new games without copying code or assets from the originals. The taxonomy organizes candidates by functional capability rather than genre, reducing coverage driven by genre popularity. We write a brief for each candidate and assign an initial task tier.

\textbf{Screening.}
Three human annotators check whether each mechanic can be evaluated with a deterministic criterion. We retain it only when correctness can be expressed as a unique value, a legal event ordering, or a quantity that must be conserved during the run. We reject mechanics whose correctness depends on visual or experiential quality. The annotators also check that the mechanic poses a meaningful implementation challenge and that its game supports complete runs, then confirm or revise its initial tier assignment.

\textbf{Agent-assisted construction.}
\leavevmode Within this construction phase, three stages turn a screened mechanic into an executable task. \emph{First,} a prototyping agent builds a small test of the mechanic's core interaction. It compares proper and naive implementations across a range of parameter settings. We keep only mechanics that produce a clear and consistent gap, and use the resulting feasibility report to guide the blueprint. \emph{Next,} a blueprint agent turns the report into a task blueprint and validates it with task runs. We discard mechanics that can be implemented directly from the brief, since they do not require reasoning about runtime interactions. \emph{Finally,} an implementation agent builds the task judge and the project shown to the evaluated agent, together with the calibration artifacts: a proper solution, a naive solution, single-capability mutants, and a behavior-preserving control. These calibration artifacts are described in Section~\ref{subsec:gate}.

\subsection{Benchmark Verification}
\label{subsec:gate}

We accept tasks after they pass calibration, agent review, and human review.

\textbf{Calibration.}
Calibration serves two purposes. First, it verifies that the task is robustly solvable across implementations. The \emph{proper solution}, a correct implementation of the required behavior, must pass every scenario and seed with sufficient margin from the tolerance boundary. A \emph{behavior-preserving control}, an alternative implementation with the same observable behavior, must also pass. Second, it verifies that the judge is sensitive to the intended capabilities. The \emph{naive solution}, an implementation with common logic errors, must fail. Each \emph{single-capability mutant}, a variant with exactly one intended capability removed, must be caught by at least one scenario designed to expose the capability it removes. Candidates that fail these checks are relaxed, hardened, or discarded. The ablation in Section~\ref{subsec:benchmark_design_ablation} measures the value of this gate.

\textbf{Agent review.}
An independent review agent performs a final verification by rerunning the calibrated task across all scenarios and seeds, checking that the judge reproduces the calibration results, and verifying the information boundary defined in Section~\ref{subsec:anatomy}. Only tasks that pass this check proceed to human review.

\textbf{Human review.}
The three annotators make the final admission decision, assessing whether tasks measure the intended mechanism, fit the benchmark's scope, and broaden capability coverage. Tasks are admitted, returned for revision, or deprecated, with tier assignments confirmed or adjusted as needed. Of roughly 200 candidate ideas, 122 completed construction, calibration, and agent review, and 72 were admitted to the library.

\subsection{Benchmark Coverage and Scale}
\label{subsec:statistics}

\begin{wraptable}{r}{\sidewrapwidth}
\vspace{-2mm}
\setlength{\abovecaptionskip}{0pt}
\caption{Library size by task tier (mean/range).}
\label{tab:benchmark_composition}
\centering
\begin{center}
\resizebox{\linewidth}{!}{%
\renewcommand{\arraystretch}{1.3}
\begin{tabular}{@{}lcccc@{}}
\toprule
\textbf{Tier} & \textbf{Tasks} & \textbf{Game files} & \textbf{Game lines} & \textbf{Proper-solution lines} \\
\midrule
Atom  & 21 & 5 (5--7)    & 316 (208--492)          & 63 (20--180)  \\
Combo & 28 & 5 (5--6)    & 421 (243--612)          & 136 (41--407) \\
Repo  & 23 & 209 (5--737) & 17{,}599 (515--62{,}895) & 316 (57--945) \\
\bottomrule
\end{tabular}%
}
\end{center}
\vspace{-4mm}
\end{wraptable}

\begin{figure}[t]
\begin{center}
\includegraphics[width=\textwidth]{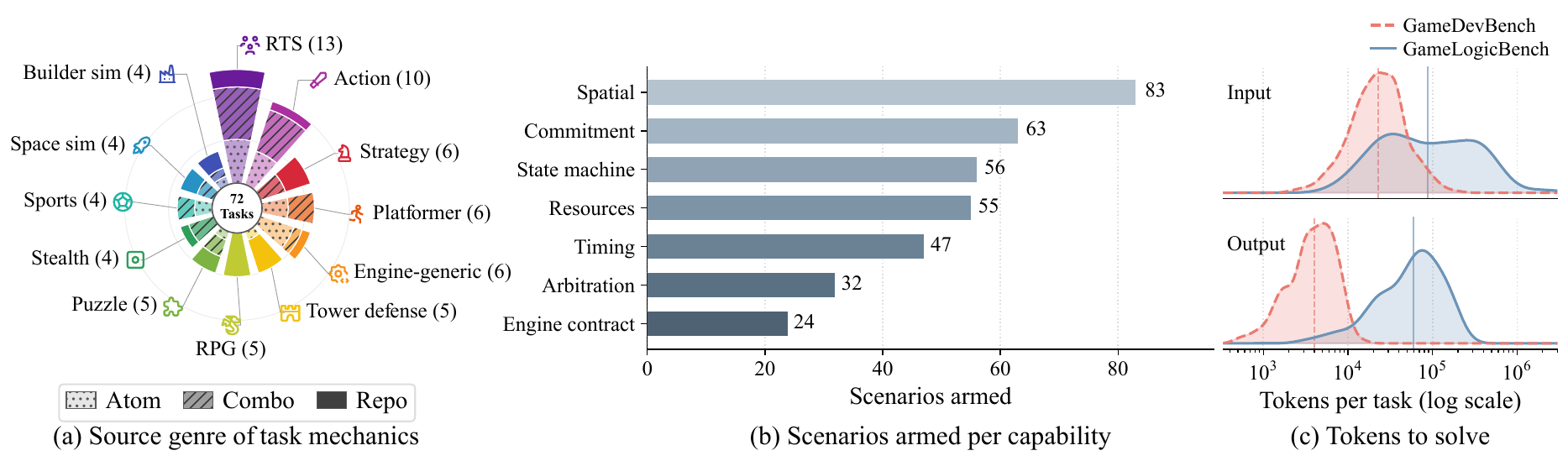}
\end{center}
\caption{Task mechanics are grouped by source genre (\textbf{left}), scenarios are grouped by capability class (\textbf{center}), and token use is compared between \benchname{} and GameDevBench (\textbf{right}).}
\label{fig:benchmark_statistics}
\end{figure}

The 72-task library spans twelve game genres. Its scored scenarios cover the seven capability classes defined in Appendix~\ref{app:capability_profiles}, with spatial navigation the most frequent. Figure~\ref{fig:benchmark_statistics} summarizes the benchmark's source composition, capability coverage, and token demand. The library contains 403 scenarios and 1,451 test cases. Each task contains 2--12 scenarios and 10--38 test cases.

Table~\ref{tab:benchmark_composition} shows an integration-scale gradient across the three tiers: Repo tasks are substantially larger than Atom and Combo tasks. Beyond static project size, a matched comparison using GLM-5.2 with OpenCode shows that \benchname{} uses about 15$\times$ more output tokens and nearly 4$\times$ more non-cached input tokens than GameDevBench~\citep{gamedevbench2026}.

\section{Evaluation}
\label{sec:eval}

\subsection{Evaluation Protocol}
\label{subsec:pipeline}

\textbf{Solving.}
The solve container provides the agent-visible task materials defined in Section~\ref{subsec:anatomy}. The agent can run the game and reseed the previewed scenario, while outbound access beyond the model API is blocked. After solving, the agent's solution is submitted for evaluation. All benchmark scores come from sealed runs. Section~\ref{subsec:hack} reports what agents attempted when a route out existed.

\textbf{Judging.}
The judge runs in a separate network-disabled container using the frozen judge files and a copy of the agent's workspace. For every scored scenario and seed, it runs the game at a fixed timestep and evaluates tick-level assertions. It varies call schedules permitted by the agent-visible interface and caller contract, including concurrent calls, re-entry, and stretched time bases. These variations change the execution conditions, not the required behavior. Its assertions inspect runtime state and event history rather than source code or implementation choices. With the runtime and judge code fixed and no language model in the scoring path, replaying the same submission, scenario, and seed produces the same verdict. A hand-designed scenario defines the test structure, and a seed instantiates its numeric parameters to produce one test case. The same assertions are applied across scenarios and seeds, so each task specification is tested under varied legal inputs.

\subsection{Experimental Setup}
\label{subsec:setup}

We evaluate twenty configurations, each pairing a model with one of three scaffolds: Claude Code 2.1.177~\citep{anthropic2026claudecode}, Codex 0.144.1~\citep{openai2026codexcli}, or OpenCode 1.17.18~\citep{anomaly2026opencode}. Both solving and judging use Godot 4.4. All solve sessions use \texttt{effort=high} and a 3,600-second time limit. Appendix~\ref{app:models} provides details of the models used in the evaluation.

\section{Results}
\label{sec:results}
\label{sec:experiments} 

\subsection{Performance and Cost}
\label{subsec:main}

\textbf{Broader integration scope remains a consistent source of difficulty.} Table~\ref{tab:main_results} reports solve rates by tier and the efficiency measures for all configurations. The best-performing configuration, Claude-Opus-5 with Claude Code, solves 52.78\% of the library. Solve rates decline from Atom to Combo to Repo for nearly every configuration.

\begin{table*}[t]
\centering
\begingroup
\definecolor{best}{RGB}{0, 140, 100}
\definecolor{second}{RGB}{30, 100, 180}
\definecolor{scaffoldlabel}{RGB}{74, 55, 96}
\renewcommand{\best}[1]{\textcolor{best}{\textbf{#1}}}
\renewcommand{\second}[1]{\textcolor{second}{\underline{#1}}}
\caption{Solve rates per tier, one sample per configuration, all runs with egress sealed. \best{Green bold} marks the best value in a ranked column and \second{blue underline} the second-best. Cost is the total over all tasks. Turns is the mean number of agent turns per task.}
\label{tab:main_results}
\fontsize{7}{8.4}\selectfont
\setlength{\tabcolsep}{3pt}
\renewcommand{\arraystretch}{1.18}
\begin{center}
\resizebox{\textwidth}{!}{%
\begin{tabular}{@{}>{\raggedright\arraybackslash}m{0.11\textwidth}>{\raggedright\arraybackslash}m{0.22\textwidth}*{6}{>{\centering\arraybackslash}m{0.095\textwidth}}@{}}
\toprule
\multicolumn{1}{c}{\multirow{2}{*}{\textbf{Scaffold}}} & \multicolumn{1}{c}{\multirow{2}{*}{\textbf{Model}}} & \multicolumn{4}{c}{\textbf{Solve rate (\%)}} & \multicolumn{2}{c}{\textbf{Efficiency}} \\
\cmidrule(lr){3-6}\cmidrule(l){7-8}
& & \textbf{Atom} & \textbf{Combo} & \textbf{Repo} & \textbf{Average} & \textbf{Turns} & \textbf{Cost (USD)} \\
\midrule
\multirow{12}{*}{\textit{\textcolor{scaffoldlabel}{Claude Code}}}
& Claude-Opus-5           & \best{61.90} & \best{53.57} & \best{43.48} & \best{52.78} & 52.5 & 443.16 \\
& Qwen-3.8-Max            & \best{61.90} & \second{50.00} & 21.74 & \second{44.44} & 57.8 & 309.91 \\
& GPT-5.6-Sol             & \second{57.14} & 35.71 & \second{34.78} & 41.67 & 37.2 & 83.97 \\
& DeepSeek-V4-Pro-0813    & 52.38 & 39.29 & \second{34.78} & 41.67 & 50.4 & 19.68 \\
& GLM-5.2                 & 52.38 & 32.14 & 30.43 & 37.50 & 40.7 & 77.56 \\
& Kimi-K3                 & 52.38 & 42.86 & 13.04 & 36.11 & 35.0 & 85.01 \\
& DeepSeek-V4-Flash-0731  & 38.10 & 32.14 & 26.09 & 31.94 & 43.4 & \best{3.59} \\
& Claude-Sonnet-5         & 47.62 & 21.43 & 8.70 & 25.00 & 37.2 & 119.50 \\
& DeepSeek-V4-Pro         & 33.33 & 28.57 & 13.04 & 25.00 & 40.5 & 8.41 \\
& MiniMax-M3              & 28.57 & 17.86 & 13.04 & 19.44 & 57.1 & 16.92 \\
& MiMo-V2.5-Pro           & 33.33 & 10.71 & 8.70 & 16.67 & 31.9 & \second{4.26} \\
& DeepSeek-V4-Flash       & 23.81 & 14.29 & 13.04 & 16.67 & 36.7 & 5.07 \\
\midrule
\multirow{4}{*}{\textit{\textcolor{scaffoldlabel}{Codex}}}
& GPT-5.6-Sol             & 47.62 & 32.14 & 26.09 & 34.72 & 12.7 & 73.16 \\
& GPT-5.6-Terra           & 42.86 & 25.00 & 21.74 & 29.17 & 9.3 & 23.70 \\
& Qwen-3.8-Max            & \second{57.14} & 21.43 & 4.35 & 26.39 & 14.1 & 90.94 \\
& Kimi-K3                 & 52.38 & 21.43 & 8.70 & 26.39 & 14.2 & 142.39 \\
\midrule
\multirow{4}{*}{\textit{\textcolor{scaffoldlabel}{OpenCode}}}
& GPT-5.6-Sol             & 42.86 & 35.71 & 26.09 & 34.72 & 14.1 & 57.61 \\
& GLM-5.2                 & \second{57.14} & 21.43 & 21.74 & 31.94 & 48.6 & 89.32 \\
& Qwen-3.8-Max            & 52.38 & 25.00 & 13.04 & 29.17 & 23.2 & 85.40 \\
& Kimi-K3                 & 33.33 & 17.86 & 17.39 & 22.22 & 43.1 & 156.60 \\
\bottomrule
\end{tabular}%
}
\end{center}
\endgroup
\end{table*}

\textbf{Performance also varies with the scaffold.} Qwen-3.8-Max's solve rate ranges from 26.39\% to 44.44\% across the three scaffolds. These differences make the model--scaffold pair, rather than the model alone, the relevant unit of comparison.

\textbf{Higher inference cost does not consistently translate into higher task success.} Configurations with the same solve rate differ by up to \(25\times\) in cost. The cost--performance frontier contains only Claude Code configurations, and its endpoints have solve rates of 31.94\% and 52.78\%, with costs differing by a factor of 123.

\subsection{Tier Effects}
\label{subsec:tier}

\begin{figure}[t]
\begin{center}
\includegraphics[width=\textwidth]{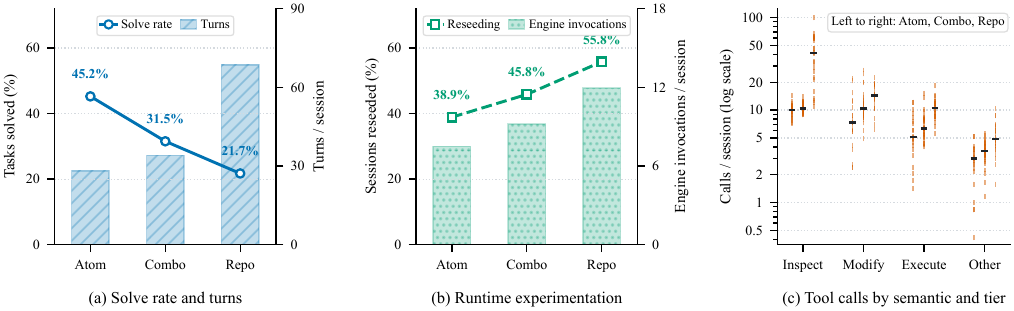}
\end{center}
\caption{Under Claude Code, solve rate and turns are reported across task tiers (\textbf{left}), reseeding and engine invocations summarize runtime experimentation (\textbf{center}), and per-task mean tool-call counts are grouped by semantic, with black bars marking group medians (\textbf{right}).}
\label{fig:tier_gradient}
\end{figure}

\textbf{Agents use executable feedback during solving.} Across all configurations and tiers, they launch the engine at least once in 99.0\% of sessions and run a Godot script written for their own experiments in 39.2\%. With Claude Code fixed, Figure~\ref{fig:tier_gradient} shows that all twelve model configurations spend more turns on Repo tasks than on Atom tasks. As task scope increases, agents launch the engine more often and reseed the previewed scenario in a larger share of sessions.

\textbf{Modification and execution calls increase from Atom to Combo, while inspection calls change little between these tiers and rise sharply on Repo tasks.} The higher inspection counts may reflect the extra work of locating relevant code and tracing dependencies in larger projects.

\subsection{Scaffold Effects}
\label{subsec:scaffold}

\begin{figure}[htbp]
\begin{center}
\includegraphics[width=\textwidth]{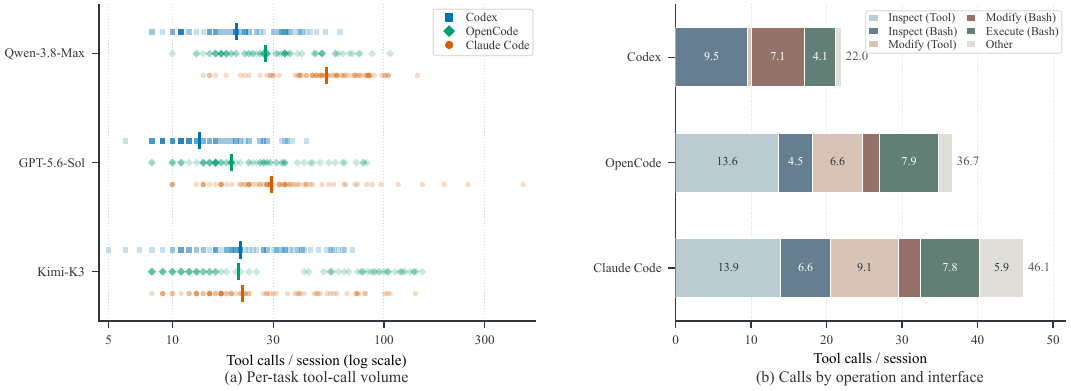}
\end{center}
\caption{Tool use varies both across tasks and scaffolds. Per-task call distributions are shown on a log axis, with medians marked by ticks (\textbf{left}); mean calls are broken down by scaffold-provided tool versus Bash and by primary operation (\textbf{right}).}
\label{fig:scaffold_tool_fingerprints}
\end{figure}

\textbf{Scaffold choice affects both solve rate and interaction volume.} The scaffold ablation in Table~\ref{tab:main_results} shows that Claude Code yields the highest solve rate for every model, while Codex uses the fewest turns. Figure~\ref{fig:scaffold_tool_fingerprints} shows that Codex also uses fewer tool calls than the other scaffolds for Qwen-3.8-Max and GPT-5.6-Sol at every reported quantile from Q1 to P90. Kimi-K3 has similar medians across scaffolds, but OpenCode produces a much longer upper tail. Appendix~\ref{app:toolcalls} provides the corresponding quantile values.

\textbf{Scaffold interfaces also induce distinct operation patterns.} Figure~\ref{fig:scaffold_tool_fingerprints}(b) shows that Codex routes all inspection and most modification calls through Bash, with its built-in patch tool handling the rest. Claude Code and OpenCode expose dedicated inspection and editing tools. The Bash audit in Appendix~\ref{app:toolcalls} shows that compound calls account for 52.5--54.7\% of Bash calls across the three scaffolds. Codex more often combines all three primary operations, whereas Claude Code and OpenCode more often combine inspection with execution.

\textbf{Interaction volume does not map to performance in the same way for every model.} Table~\ref{tab:main_results} shows that Qwen-3.8-Max reaches its highest solve rate under Claude Code, which is also its most expensive scaffold. It uses the most tool calls and input tokens in that setting, as reported in Appendices~\ref{app:toolcalls} and~\ref{app:tokens}. Kimi-K3 shows the reverse cost pattern. Claude Code is its least expensive and highest-scoring scaffold. OpenCode produces Kimi-K3's largest call volume but its lowest solve rate.

\subsection{Attempt Stability}
\label{subsec:stability}

We measure how consistently each configuration solves tasks across repeated attempts. For the three configurations evaluated three times, Table~\ref{tab:sampling_variance} reports four metrics, where pass@1 is the mean task solve rate across attempts, pass@3 is the fraction of tasks solved at least once, worst@3 is the fraction solved in all three attempts, and $\sigma$ is the standard deviation of the three single-attempt scores.

\begin{wraptable}{r}{\sidewrapwidth}
\vspace{-4mm}
\caption{Task solve rates and variation.}
\label{tab:sampling_variance}
\begin{center}
\resizebox{\linewidth}{!}{%
\renewcommand{\arraystretch}{1.3}
\begin{tabular}{@{}lcccc@{}}
\toprule
\textbf{Model} & \textbf{pass@1 (\%)} & \textbf{pass@3 (\%)} & \textbf{worst@3 (\%)} & $\boldsymbol{\sigma}$ \textbf{(\%)} \\
\midrule
Qwen-3.8-Max & 45.37 & 62.50 & 27.78 & 4.24 \\
GLM-5.2      & 34.72 & 48.61 & 22.22 & 3.67 \\
Kimi-K3      & 34.72 & 51.39 & 15.28 & 3.67 \\
\bottomrule
\end{tabular}%
}
\end{center}
\vspace{-4mm}
\end{wraptable}

\textbf{The set of solved tasks varies across repeated attempts.} The pass@3--worst@3 gap is 26.39\%--36.11\% across the three configurations. Within configurations, the standard deviation of the three single-attempt scores ranges from 3.67\% to 4.24\%, equivalent to two or three tasks.

\subsection{Capability-Conditioned Performance}
\label{subsec:profile}
\label{subsec:concentration}

\textbf{Maintaining task-specific decisions and spatial and temporal invariants is harder than basic engine use.} Across the 20 configurations, mechanism failures account for 74.3\% of failed scenarios, meaning that a submitted solution runs and is judgeable but violates a behavioral contract, while 17.2\% have no judgeable solution and 8.5\% are nonviable computations. To locate where the behavioral requirement is hardest to maintain, we use the pre-specified capability classes and count a strict scenario pass only when every seed passes. Figure~\ref{fig:failure_attribution_full} shows that Engine contract ranks highest among the seven classes in 19 of 20 configurations, with a pooled strict pass rate of 88.75\%. In contrast, Commitment, Spatial, and Timing are the bottom three classes in 17 of 20 configurations, with pooled strict pass rates between 53.4\% and 58.57\%. This 30\%--35\% separation indicates that basic use of Godot's clock, solver, and time base is usually not the limiting step.

\subsection{Benchmark Design Ablations}
\label{subsec:benchmark_design_ablation}

\begin{wraptable}{r}{\sidewrapwidth}
\vspace{-4mm}
\caption{Effects of evaluation ablations relative to the full evaluator.}
\label{tab:evaluation_design_ablation}
\begin{center}
\resizebox{\linewidth}{!}{%
\renewcommand{\arraystretch}{1.3}
\begin{tabular}{@{}lcc@{}}
\toprule
\textbf{Metric} & \textbf{Terminal-only} & \textbf{Preview-only} \\
\midrule
Escaped mutants & 236 / 666 (35.4\%) & 508 / 666 (76.3\%) \\
Tasks with escapes & 34 / 36 & 36 / 36 \\
FAIL$\rightarrow$PASS solutions & 64 / 488 (13.1\%) & 418 / 488 (85.7\%) \\
Mean $\Delta$ solve rate & +8.9\% & +58.1\% \\
\bottomrule
\end{tabular}%
}
\vspace{-6mm}
\end{center}
\end{wraptable}

We test two layers of benchmark design on the same 36-task audit set: the scoring protocol and the criterion-validation gate.

\textbf{Scoring protocol.} We vary the evidence available to the evaluator while holding the tasks and submitted solutions fixed. We rescore the same mutants and every submitted solution from the 20 main-table configurations under two controlled ablations.
Terminal-only retains all scenarios and seeds but uses only terminal-state assertions, whereas Preview-only retains the full tick-level criterion but evaluates only the preview scenario.

Table~\ref{tab:evaluation_design_ablation} shows why both tick-level assertions and judge-selected scenarios are needed, since removing either one allows more invalid behavior to pass. Terminal-only lets mutants escape on 34 of 36 tasks despite retaining all scenarios and seeds, demonstrating that terminal-state assertions miss violations during execution. Preview-only is even less discriminating, allowing mutants to escape on all 36 tasks and many solutions that fail the full evaluator to pass the visible scenario.

\textbf{Criterion validation.} During benchmark construction, the tasks' original criteria without mutants serve as the control. We then add mutant validation and repair the gaps it exposes. Correct reference solutions test whether a judge accepts valid behavior, whereas mutants test whether it rejects implementations that deliberately violate one requirement.

In the control, 127 of 666 mutants passed, revealing 24 missing checks across 19 of the 36 tasks. We compared task outcomes before and after repair for 9 models under 2 scaffolds. The repair changed 3 task outcomes from PASS to FAIL, affecting 2 tasks and 3 models. Proper solutions and behavior-preserving controls still passed. Validating only correct solutions therefore leaves criterion gaps large enough to change benchmark outcomes.

\FloatBarrier
\subsection{Retrieval Contamination Audit}
\label{subsec:hack}

\begin{figure}[htbp]
\begin{center}
\includegraphics[width=\textwidth]{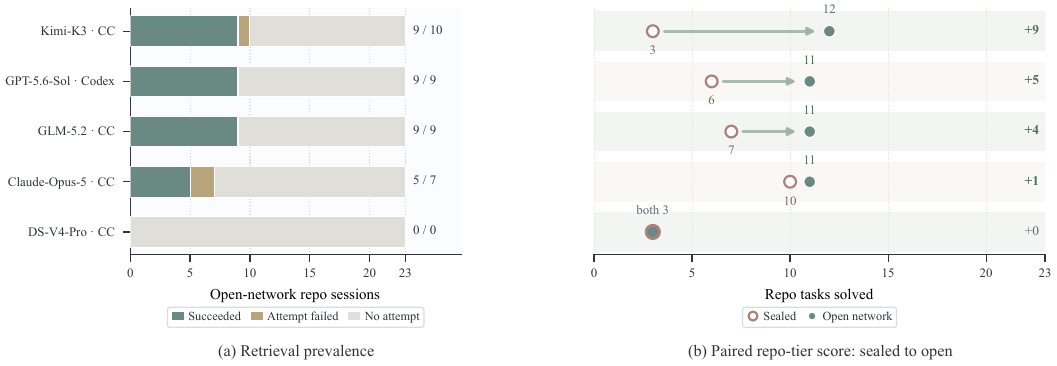}
\end{center}
\caption{Open-network repository sessions are classified by retrieval outcome (\textbf{left}). The right panel compares each configuration's repo-tier score under sealed and open-network access (\textbf{right}).}
\label{fig:upstream_retrieval}
\end{figure}

\textbf{Repository-derived tasks are vulnerable to upstream source retrieval when the network is open.} We evaluated the Repo tier under open-network access for five configurations, pairing each run with its sealed counterpart. We scanned solve trajectories for URLs, queries, and shell commands, then inspected the returned content. Four configurations retrieved upstream source, and manual inspection found direct reuse in the reviewed traces. Figure~\ref{fig:upstream_retrieval} shows that all four configurations achieved higher Repo-tier solve rates in the open-network condition than in the sealed condition.

\FloatBarrier

\section{Related Work}
\label{sec:related}


\textbf{Benchmarks for coding agents.}
Executable tests have graded code generation since HumanEval and MBPP posed isolated functions against fixed unit tests~\citep{humaneval2021,mbpp2021}. SWE-bench scaled that form to real repositories by pairing GitHub issues with the tests their accepted patches satisfy~\citep{swebench2024}, SWE-bench Multimodal carried it into visual software domains~\citep{swebenchmm2024}, and Terminal-Bench collects hard, expert-authored tasks in the command line~\citep{terminalbench2026}. For interactive artifacts, WebCompass uses checklist-guided model judges and an agent judge that drives the artifact in a real browser~\citep{webcompass2026}, and PlayCoder sends a vision-language tester to play generated graphical applications~\citep{playcoder2026}. Our judge keeps the deterministic verdict this lineage began with and follows the artifact into its run.

\textbf{Game-development benchmarks.}
Benchmarks for complete game generation evaluate whether an artifact implements the requested mechanics and delivers a playable experience. To assess implementation quality and presentation, V-GameGym, OpenGame-Bench, and GameCraft-Bench use model judgments of generated code or rendered gameplay~\citep{vgamegym2026,opengame2026,gamecraft2026}. Interactive evaluation checks requirements through agent playtesting, as in WebGameBench, PlaytestArena, and AutoUE~\citep{webgamebench2026,playtestarena2026,autoue2026}.
Benchmarks built around game-engine projects examine how generated code integrates with scene structure and runtime behavior. Evaluations in Godot and Unity use engine test scripts, reference-based runtime measurements, and static mechanism analysis to assess different aspects of generated projects~\citep{gamedevbench2026,jamer2026,mage2026}. Runtime testing also verifies timed mechanics in OpenGameEval's Roblox tasks, including delayed health regeneration~\citep{opengameeval2025}. GameEngineBench evaluates C++ changes in running Unreal projects, using behavioral tests and an LLM judge to determine whether the requested behavior is satisfied~\citep{gameenginebench2026}. \benchname complements these efforts by focusing on gameplay-logic correctness throughout execution, with deterministic tick-level assertions for every task across multiple judge-selected scenarios. Table~\ref{tab:benchmark_comparison} summarizes selected evaluation designs.

\section{Conclusion}
\label{sec:conclusion}

We introduce \benchname, a benchmark of 72 gameplay-logic tasks in Godot projects. Its deterministic judge checks tick-level invariants across multiple judge-selected scenarios. Across 20 model--scaffold configurations, the best result solves 52.78\% of the task library. In 74.3\% of failed scenarios, the solution is still runnable but violates required behavior. Our evaluation reveals that the main challenge is maintaining correct behavior during execution, not merely producing a runnable project. The scaffold ablation shows that performance and tool use depend on the model--scaffold pairing rather than the model alone. Separately, the benchmark-design ablations identify two complementary requirements for reliable scoring. First, the scoring protocol must combine tick-level assertions with judge-selected scenarios. Second, criterion validation must also use mutants.

\section*{Limitations and Future Work}
\label{sec:limitations}

\benchname focuses on gameplay logic that can be checked with deterministic, tick-level state assertions. It does not evaluate content depth, art, presentation, or overall player experience, which are addressed by benchmarks of complete games~\citep{gamecraft2026,webgamebench2026}. Future versions could cover longer interactions and additional runtime signals while keeping verdicts deterministic.

Repo coverage draws on permissively licensed, runnable open-source Godot projects, so its genre mix reflects the available upstream repositories. The benchmark is also limited to Godot and GDScript and does not test network synchronization within its single-container harness. Future releases could broaden genre and engine coverage through targeted sourcing and add networked evaluation through multi-container client--server setups.

\section*{Ethical Considerations}
\label{sec:ethics}

Twenty Repo tasks are derived from public open-source Godot projects.
We redistribute only materials with compatible licenses, retain the required
copyright and attribution notices, and document the sources and our changes.
Assets that cannot be redistributed are replaced or omitted.

\section*{AI Use Statement}

We used generative AI agents, including Fable 5 and Claude Opus 4.8, in the
design and construction of the benchmark tasks. Human reviewers also reviewed
the benchmark tasks during construction. We used Fable 5 and GPT-5.6-Sol to
help polish the exposition of the paper.

\bibliographystyle{unsrtnat}
\bibliography{references}

\clearpage
\appendix

\section{Model Card}
\label{app:models}

Table~\ref{tab:model_card} lists the model releases and vendor prices used to compute the Cost column of Table~\ref{tab:main_results}. We reverified these prices on 2026-08-15. For each DeepSeek line, the undated name denotes the original April 2026 release and the dated suffix denotes a newer snapshot. Both snapshots shared one list price during evaluation. DeepSeek's subsequent repricing took effect after the experiments ended.

The corresponding cost--performance view is shown in Figure~\ref{fig:main_results_cost}. It places every model--scaffold configuration at its observed solve rate and total inference cost; the marked frontier is descriptive rather than a new ranking criterion.

\begin{table}[htbp]
\caption{Vendor prices in USD per 1M tokens. Anthropic cache writes use the 5-minute retention tier; a dash denotes no separate cache-write charge.}
\label{tab:model_card}
\begin{center}
\resizebox{\textwidth}{!}{%
\renewcommand{\arraystretch}{1.3}
\begin{tabular}{@{}llrrrr@{}}
\toprule
\textbf{Model} & \textbf{Released} & \multicolumn{4}{c}{\textbf{Price (USD / 1M tokens)}} \\
 & & \textbf{Input} & \textbf{Cache write} & \textbf{Cache read} & \textbf{Output} \\
\midrule
\texttt{claude-opus-5} \citep{anthropic2026opus5} & 2026-07-24 & 5.00 & 6.25 & 0.50 & 25.00 \\
\texttt{claude-sonnet-5} \citep{anthropic2026sonnet5} & 2026-06-30 & 2.00 & 2.50 & 0.20 & 10.00 \\
\texttt{deepseek-v4-flash} \citep{deepseek2026v4preview} & 2026-04-24 & 0.14 & --- & 0.0028 & 0.28 \\
\texttt{deepseek-v4-flash-0731} \citep{deepseek2026flash0731} & 2026-07-31 & 0.14 & --- & 0.0028 & 0.28 \\
\texttt{deepseek-v4-pro} \citep{deepseek2026v4preview} & 2026-04-24 & 0.435 & --- & 0.003625 & 0.87 \\
\texttt{deepseek-v4-pro-0813} \citep{deepseek2026pro0813} & 2026-08-13 & 0.435 & --- & 0.003625 & 0.87 \\
\texttt{glm-5.2} \citep{zai2026glm52} & 2026-06-16 & 1.40 & --- & 0.26 & 4.40 \\
\texttt{gpt-5.6-sol} \citep{openai2026gpt56} & 2026-07-09 & 5.00 & --- & 0.50 & 30.00 \\
\texttt{gpt-5.6-terra} \citep{openai2026gpt56} & 2026-07-09 & 2.00 & --- & 0.20 & 12.00 \\
\texttt{kimi-k3} \citep{moonshot2026kimik3} & 2026-07-16 & 3.00 & --- & 0.30 & 15.00 \\
\texttt{mimo-v2.5-pro} \citep{xiaomi2026mimov25pro} & 2026-04-27 & 0.435 & --- & 0.0036 & 0.87 \\
\texttt{minimax-m3} \citep{minimax2026m3} & 2026-06-01 & 0.30 & --- & 0.06 & 1.20 \\
\texttt{qwen3.8-max} \citep{qwen2026max38} & 2026-08-03 & 2.00 & --- & 0.25 & 6.00 \\
\bottomrule
\end{tabular}%
}
\end{center}
\end{table}

\FloatBarrier
\section{Capability Classes and Per-Model Profiles}
\label{app:capability_profiles}

\begin{figure}[htbp]
\begin{center}
\includegraphics[width=\textwidth]{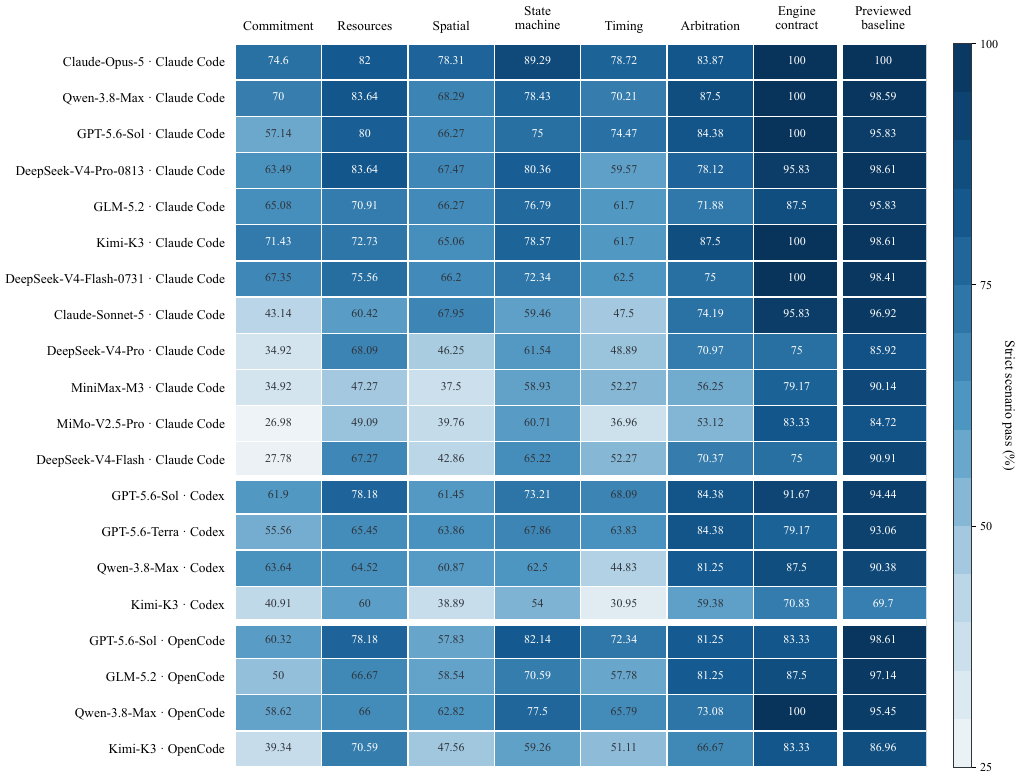}
\end{center}
\caption{Capability-conditioned strict pass rates for all 20 configurations.}
\label{fig:failure_attribution_full}
\end{figure}

Table~\ref{tab:capability_vocabulary} defines the seven capability classes used in Section~\ref{subsec:profile}. A scenario may belong to more than one class. Figure~\ref{fig:failure_attribution_full} reports strict scenario pass rates for all 20 configurations, with the previewed baseline shown separately.

\begin{table}[htbp]
\caption{Definitions of the seven capability classes.}
\label{tab:capability_vocabulary}
\begin{center}
\small
\begin{tabular}{@{}>{\raggedright\arraybackslash}p{0.23\textwidth}>{\raggedright\arraybackslash}p{0.71\textwidth}@{}}
\toprule
\textbf{Capability class} & \textbf{Behavior covered} \\
\midrule
Intention commitment
  & Multi-frame commitment, hysteresis, resistance to dithering, and decisions that must remain stable while their consequences unfold. \\
Resource accounting
  & Conservation, allocation from shared scarce pools, bookkeeping, and propagation through supply networks. \\
Spatial navigation and steering
  & Position, orientation, reachability, replanning, avoidance, interception, and contact classification. \\
State machine
  & State transitions, lifecycle changes, inherited state after a transition, termination conditions, and gates. \\
Timing
  & Frame- or second-scale windows, cooldowns, cadence, phase alignment, and temporal invariants. \\
Arbitration
  & Priority rules, same-frame adjudication, yielding, deduplication, and selection among competing claims. \\
Engine contract
  & Correct use of the engine's clock, solver, time base, and subsystem semantics. \\
\bottomrule
\end{tabular}
\end{center}
\end{table}

\FloatBarrier
\section{Per-Task Coverage}
\label{app:coverage}

\begin{figure}[htbp]
\begin{center}
\includegraphics[width=\textwidth,height=0.70\textheight,keepaspectratio]{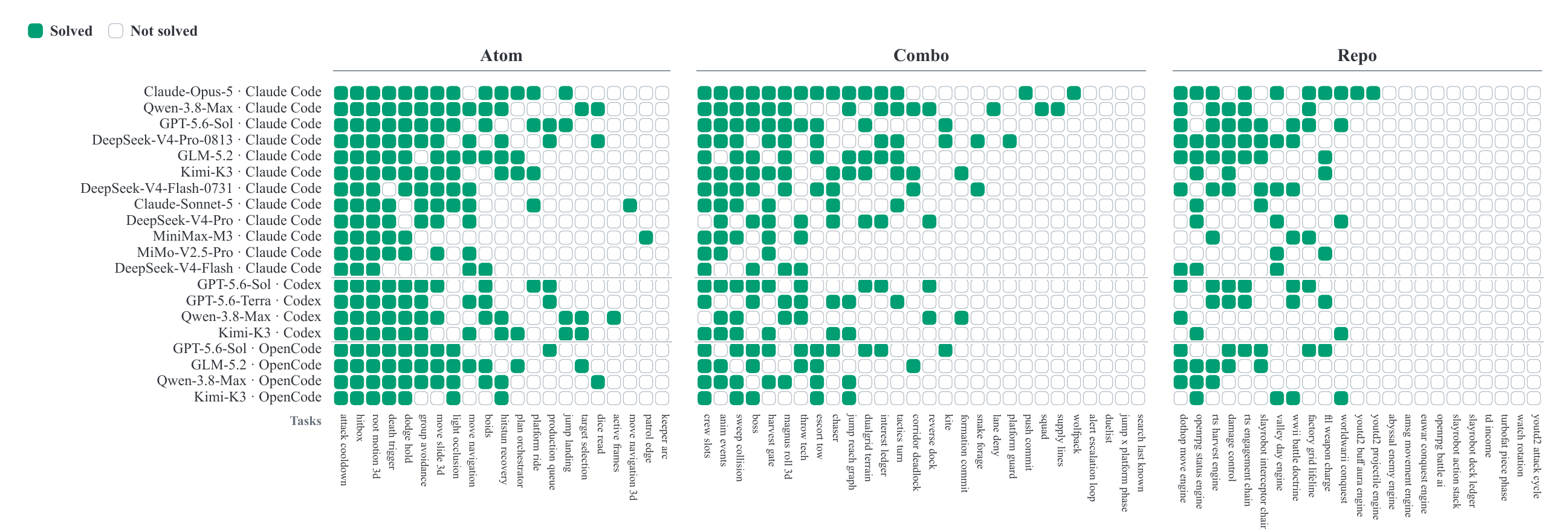}
\end{center}
\caption{Per-task outcomes by configuration and tier.}
\label{fig:coverage_grid}
\end{figure}

Figure~\ref{fig:coverage_grid} expands the tier-level solve rates in Table~\ref{tab:main_results} into per-task outcomes for all 20 configurations. Within each tier, tasks are ordered by solve count.

\FloatBarrier
\section{Tool-Call Statistics}
\label{app:toolcalls}

Table~\ref{tab:scaffold_tool_profile} reports mean tool use across all 20 configurations. The four operation shares are mutually exclusive; Bash share is reported separately, and Bash failure is computed over Bash calls. Table~\ref{tab:scaffold_call_distribution} reports the call-count quantiles underlying Figure~\ref{fig:scaffold_tool_fingerprints}(a) for the three models evaluated under every interface. Table~\ref{tab:bash_operation_audit} decomposes Bash calls into primary and co-occurring operations. Each call receives one primary label under the priority modify $>$ execute $>$ inspect $>$ other. Facet labels may co-occur, while the combination rows form a disjoint partition.

\begin{table}[htbp]
\caption{Mean tool-use profiles by configuration.}
\label{tab:scaffold_tool_profile}
\begin{center}
\resizebox{\textwidth}{!}{%
\begin{tabular}{@{}llrrrrrrr@{}}
\toprule
\textbf{Scaffold} & \textbf{Model} & \textbf{Calls/session}
& \textbf{Inspect (\%)} & \textbf{Modify (\%)} & \textbf{Execute (\%)}
& \textbf{Other (\%)} & \textbf{Bash (\%)} & \textbf{Bash failure (\%)} \\
\midrule
\multirow{12}{*}{Claude Code}
  & Claude-Opus-5             & 51.8 & 35.0 & 42.9 & 21.8 & 0.3 & 51.0 & 1.1 \\
  & Qwen-3.8-Max              & 56.9 & 36.0 & 32.7 & 19.7 & 11.6 & 37.8 & 1.2 \\
  & GPT-5.6-Sol               & 47.3 & 54.4 & 14.6 & 10.9 & 20.1 & 30.2 & 4.3 \\
  & DeepSeek-V4-Pro-0813      & 49.5 & 46.5 & 31.7 & 20.8 & 0.9 & 44.4 & 1.1 \\
  & GLM-5.2                   & 39.8 & 46.3 & 30.6 & 21.0 & 2.1 & 42.5 & 1.7 \\
  & Kimi-K3                   & 34.1 & 44.6 & 30.4 & 20.3 & 4.8 & 46.2 & 1.0 \\
  & DeepSeek-V4-Flash-0731    & 42.4 & 47.8 & 27.3 & 19.6 & 5.3 & 37.2 & 1.4 \\
  & Claude-Sonnet-5           & 36.6 & 65.6 & 21.7 & 12.2 & 0.6 & 59.6 & 1.3 \\
  & DeepSeek-V4-Pro           & 39.6 & 47.6 & 19.2 & 15.7 & 17.5 & 30.5 & 3.1 \\
  & MiniMax-M3                & 56.1 & 39.3 & 23.5 & 18.4 & 18.8 & 34.0 & 0.8 \\
  & MiMo-V2.5-Pro             & 30.9 & 55.5 & 19.6 & 18.4 & 6.5 & 34.5 & 5.2 \\
  & DeepSeek-V4-Flash         & 36.4 & 48.2 & 21.5 & 15.4 & 14.9 & 32.4 & 4.1 \\
\midrule
\multirow{4}{*}{Codex}
  & GPT-5.6-Sol               & 16.4 & 34.3 & 42.5 & 17.0 & 6.3 & 92.7 & 15.6 \\
  & GPT-5.6-Terra             & 12.6 & 30.5 & 38.8 & 22.7 & 7.9 & 92.0 & 17.4 \\
  & Qwen-3.8-Max              & 23.2 & 52.6 & 27.0 & 17.8 & 2.6 & 97.4 & 5.1 \\
  & Kimi-K3                   & 26.4 & 40.4 & 36.0 & 20.9 & 2.7 & 92.7 & 5.5 \\
\midrule
\multirow{4}{*}{OpenCode}
  & GPT-5.6-Sol               & 27.0 & 64.5 & 0.0 & 16.7 & 18.9 & 21.3 & 15.0 \\
  & GLM-5.2                   & 53.3 & 50.4 & 25.0 & 22.7 & 2.0 & 41.6 & 2.0 \\
  & Qwen-3.8-Max              & 33.4 & 56.7 & 23.8 & 18.4 & 1.1 & 39.2 & 2.4 \\
  & Kimi-K3                   & 49.8 & 36.5 & 37.0 & 26.3 & 0.2 & 50.3 & 2.9 \\
\bottomrule
\end{tabular}%
}
\end{center}
\end{table}

\begin{table}[htbp]
\caption{Per-session tool-call quantiles in the scaffold ablation.}
\label{tab:scaffold_call_distribution}
\begin{center}
\small
\begin{tabular}{@{}llrrrr@{}}
\toprule
\textbf{Scaffold} & \textbf{Model} & \textbf{Q1} & \textbf{Median} & \textbf{Q3} & \textbf{P90} \\
\midrule
\multirow{3}{*}{Claude Code}
  & Qwen-3.8-Max & 40.0 & 53.5 & 74.5 & 86.0 \\
  & GPT-5.6-Sol  & 22.0 & 29.5 & 43.0 & 82.2 \\
  & Kimi-K3      & 15.0 & 21.5 & 47.3 & 74.9 \\
\midrule
\multirow{3}{*}{Codex}
  & Qwen-3.8-Max & 14.8 & 20.0 & 32.3 & 39.9 \\
  & GPT-5.6-Sol  & 10.8 & 13.5 & 20.0 & 27.9 \\
  & Kimi-K3      & 14.8 & 21.0 & 34.0 & 53.5 \\
\midrule
\multirow{3}{*}{OpenCode}
  & Qwen-3.8-Max & 18.0 & 27.5 & 39.5 & 60.8 \\
  & GPT-5.6-Sol  & 15.8 & 19.0 & 31.5 & 54.5 \\
  & Kimi-K3      & 12.0 & 20.5 & 87.8 & 120.8 \\
\bottomrule
\end{tabular}
\end{center}
\end{table}

\begin{table}[htbp]
\caption{Bash-operation counts by scaffold.}
\label{tab:bash_operation_audit}
\begin{center}
\small
\begin{tabular}{@{}lrrr@{}}
\toprule
\textbf{Bash classification} & \textbf{Claude Code} & \textbf{Codex} & \textbf{OpenCode} \\
\midrule
\multicolumn{4}{@{}l}{\emph{Primary label}} \\
Inspect & 1,415 & 2,053 & 976 \\
Execute & 1,675 & 896 & 1,709 \\
Modify  & 625 & 1,535 & 474 \\
Other   & 2 & 2 & 2 \\
\midrule
\multicolumn{4}{@{}l}{\emph{Facet present}} \\
Inspect & 3,341 & 4,327 & 2,691 \\
Execute & 2,148 & 2,033 & 2,081 \\
Modify  & 625 & 1,535 & 474 \\
\midrule
\multicolumn{4}{@{}l}{\emph{Mutually exclusive combination}} \\
Inspect only                    & 1,415 & 2,053 & 976 \\
Execute only                    & 292 & 58 & 424 \\
Modify only                     & 55 & 11 & 31 \\
Execute + inspect               & 1,383 & 838 & 1,285 \\
Inspect + modify                & 97 & 387 & 71 \\
Execute + modify                & 27 & 88 & 13 \\
Execute + inspect + modify      & 446 & 1,049 & 359 \\
No matched facet                & 2 & 2 & 2 \\
\midrule
Multi-operation share           & 52.5\% & 52.7\% & 54.7\% \\
\bottomrule
\end{tabular}
\end{center}
\end{table}

\section{Token-Use Statistics}
\label{app:tokens}

Table~\ref{tab:token_usage_profile} reports mean token use for all configurations. Values are in thousands of tokens per task; new input includes cache writes, and reuse is the cache-read share of total input. Cache reuse ranges from 52.9\% to 97.4\%, while mean turns per task range from 9.3 to 57.8. Table~\ref{tab:claude_cache_profile} separately reports cache writes and reads for Claude-Opus-5 and Claude-Sonnet-5.

\begin{table}[htbp]
\caption{Mean token use per task by configuration.}
\label{tab:token_usage_profile}
\begin{center}
\resizebox{\textwidth}{!}{%
\begin{tabular}{@{}llrrrrr@{}}
\toprule
\textbf{Scaffold} & \textbf{Model} & \textbf{New input (k/task)}
& \textbf{Cache read (k/task)} & \textbf{Output (k/task)}
& \textbf{Reuse (\%)} & \textbf{Turns/task} \\
\midrule
\multirow{12}{*}{Claude Code}
  & Claude-Opus-5             & 280.3   & 5,121.8 & 74.0  & 94.8 & 52.5 \\
  & Qwen-3.8-Max              & 1,566.2 & 2,787.7 & 79.2  & 64.0 & 57.8 \\
  & GPT-5.6-Sol               & 93.7    & 915.7   & 8.0   & 90.7 & 37.2 \\
  & DeepSeek-V4-Pro-0813      & 477.8   & 3,646.0 & 60.0  & 88.4 & 50.4 \\
  & GLM-5.2                   & 101.7   & 2,718.8 & 51.8  & 96.4 & 40.7 \\
  & Kimi-K3                   & 65.4    & 1,897.6 & 27.7  & 96.7 & 35.0 \\
  & DeepSeek-V4-Flash-0731    & 162.4   & 2,816.8 & 68.5  & 94.5 & 43.4 \\
  & Claude-Sonnet-5           & 186.9   & 2,417.6 & 71.0  & 92.8 & 37.2 \\
  & DeepSeek-V4-Pro           & 204.5   & 1,606.9 & 25.3  & 88.7 & 40.5 \\
  & MiniMax-M3                & 75.8    & 2,828.0 & 35.5  & 97.4 & 57.1 \\
  & MiMo-V2.5-Pro             & 52.0    & 1,379.2 & 36.3  & 96.4 & 31.9 \\
  & DeepSeek-V4-Flash         & 429.7   & 1,081.6 & 25.8  & 71.6 & 36.7 \\
\midrule
\multirow{4}{*}{Codex}
  & GPT-5.6-Sol               & 42.6  & 784.4 & 13.7  & 94.8 & 12.7 \\
  & GPT-5.6-Terra             & 59.6  & 487.1 & 9.4   & 89.1 & 9.3  \\
  & Qwen-3.8-Max              & 213.1 & 521.4 & 117.8 & 71.0 & 14.1 \\
  & Kimi-K3                   & 453.1 & 507.8 & 31.1  & 52.9 & 14.2 \\
\midrule
\multirow{4}{*}{OpenCode}
  & GPT-5.6-Sol               & 88.1  & 330.3   & 6.5  & 79.0 & 14.1 \\
  & GLM-5.2                   & 181.9 & 2,586.9 & 71.2 & 93.4 & 48.6 \\
  & Qwen-3.8-Max              & 234.4 & 593.9   & 94.8 & 71.7 & 23.2 \\
  & Kimi-K3                   & 446.3 & 1,649.2 & 22.8 & 78.7 & 43.1 \\
\bottomrule
\end{tabular}%
}
\end{center}
\end{table}

\begin{table}[htbp]
\caption{Mean cache traffic per task under Claude Code.}
\label{tab:claude_cache_profile}
\begin{center}
\small
\begin{tabular}{@{}lrr@{}}
\toprule
\textbf{Model} & \textbf{Cache write (k/task)} & \textbf{Cache read (k/task)} \\
\midrule
Claude-Opus-5   & 274.2 & 5,121.8 \\
Claude-Sonnet-5 & 184.6 & 2,417.6 \\
\bottomrule
\end{tabular}
\end{center}
\end{table}

\FloatBarrier
\section{Solve Instruction and Representative Task Briefs}
\label{app:task_briefs}

The evaluated agent receives a shared solve instruction and a task-specific README. We reproduce the shared instruction and one representative brief from each tier.

\subsection{Solve instruction}
\label{app:prompts:solve}

The solve instruction is shared across the library and contains no task-specific requirements. All three scaffolds receive the same text. The closing network-access note appears only in sealed runs.

\begin{promptbox}{Shared Solve Instruction}
You are developing in the Godot 4.4 game project mounted at {ws}. Your working directory is that
project root ({ws}).

Read {ws}/README.md first - it states what the game needs, exactly which file you implement
({deliverable}), the information your code receives each frame, and where your
deliverable ends.

Then implement the behavior the README asks for. You may split your logic across helper scripts
under {helper_dir} and preload them from {deliverable_base}.

You can run the project headless to try your work, e.g.:
    godot --headless --path {ws} {entry_scene}
Its [preview] lines report what happened, including any rule violations. Use them to debug.

When done, make sure your final code is saved in {ws}/{deliverable_rel} (plus any helpers it uses)
and that it delivers everything the README requires.

Note: this environment has no internet access. Outbound connections (git clone, curl, web fetch,
web search) fail immediately. Work only from the files in {ws}.
\end{promptbox}

\subsection{Atom task}

This example shows an Atom task that isolates one mechanic.

\begin{promptbox}{Task Brief --- Enemy Navigation (Atom)}
# Enemy navigation task

You are working in a small Godot 4.4 game project. An enemy must travel across a walled arena to
reach its goal. Your job is to write the enemy's movement decision so it gets there reliably.

The game builds the arena procedurally: walls, doorways, start and goal are laid out differently
from one play to the next. The preview is wired to one example arena — your enemy has to reach
its goal in whichever arena the game builds.

Press **F5** (or `godot --path . res://main.tscn`) to watch the current enemy attempt the course
and to debug your work. The preview prints what happened (`arrived`, `CLIPPED a wall`, timeout).

## Goal

Drive the enemy from its start position to the goal:

- **Reach the goal** before the time budget runs out.
- **Never let the enemy's body touch a wall.** The enemy is a circle of a given radius; grazing a
  wall corner counts as a collision and fails the run.

## Where your work goes

Implement the decision in **`res://logic/controller.gd`**:

```gdscript
func decide(state: Dictionary) -> Vector2:
    # return the DIRECTION to move this physics frame (any non-zero Vector2; it is normalized and
    # the enemy advances a fixed distance along it). Return Vector2.ZERO to hold.
```

Optional one-time setup:

```gdscript
func setup(state: Dictionary) -> void:
    # runs once before the first frame
```

You may split your logic across several scripts under `res://logic/` and `preload` them from
`controller.gd`. How you compute the direction is entirely up to you.

### What `state` gives you (world units)

| key | type | meaning |
|---|---|---|
| `self_pos` | `Vector2` | the enemy's current position |
| `goal_pos` | `Vector2` | the goal position |
| `radius`   | `float`   | the enemy's collision radius |
| `world`    | `Node2D`  | a scene handle you may use to query the physics world |
| `nav_map`  | `RID`     | a navigation map handle for the arena you may query for routing |
| `dt`       | `float`   | this frame's timestep |
| `t`        | `float`   | elapsed time |

You may use any, all, or none of these. The contract fixes only `decide()`'s signature.

## What is fixed

Your deliverable is **`res://logic/controller.gd`** plus any helper scripts it pulls in from
`res://logic/`. The rest of the project — the arena (`level.gd`), the world setup (`world_runtime.gd`),
the enemy body and driver (`enemy.gd`, `sim_core.gd`), the geometry probes (`assertions.gd`) and the
project configuration — is the game itself: your AI has to work with it exactly as it stands here.
While developing you may change anything locally — add prints, tweak the world, set up whatever
experiment helps you debug — but changes outside `res://logic/` are debugging aids, not part of your
deliverable.
\end{promptbox}

\subsection{Combo task}

This example shows a Combo task that combines several interacting mechanics.

\begin{promptbox}{Task Brief --- Platform Guard (Combo)}
# combo_platform_guard — Game Preview

Press **F5** to run the preview. You are the yard's guard: a platformer body (real
gravity, real jumps) standing watch on the home platform while visitors roam the far
side of the yard. Your `logic/controller.gd` controller drives it.

The game builds the yard procedurally: platform sizes, the gap between them, visit
timing and your spawn point are laid out differently from one play to the next. Yards
can also differ structurally — shorter home platforms, raised far decks across wider
gaps, and towers standing mid-platform that block the view. The preview is wired to
one example yard.

## Controller interface

```gdscript
func decide(state: Dictionary) -> Dictionary
```

Return `{"move": float, "jump": bool, "chasing": int}` each frame.

| field | meaning |
|-------|---------|
| `move` | -1.0 (left) … 1.0 (right); clamped to [-1,1] |
| `jump` | jump intent — only accepted while standing on ground (mid-air intent is ignored) |
| `chasing` | id of the intruder you are pursuing, or -1 when not chasing |

Gravity is real: one step past a platform edge and the guard is airborne with no way
back, and there is no floor below the platforms. A jump's launch point decides where
the arc lands — crossing a gap from too far back means falling into the pit.

## The guard's duties

- **Patrol**: through quiet stretches, sweep the home platform — cover at least **34%**
  of its walkable extent. Standing in one spot is not guarding.
- **Watch**: an intruder is *visible* when it is within `vision_range` **and** the
  sight line to it is unobstructed. Towers (and platform decks) block sight. Distance
  alone is not visibility.
- **Confront**: when a visitor is plainly visible, move in — close to within **130**
  units of it, promptly, **while standing on the ground**. Being within 130 of it in
  mid-air does not count: a pursuit jump has to *land* next to the visitor, so an arc
  that sails past it and coasts back has not confronted anyone. If it leaves or you lose
  sight of it, the confrontation is over.
- **Be honest about the chase**: set `chasing` only on a target you can actually see.
  Claiming a chase on something out of sight is a false alarm.
- **Return**: when the yard is quiet, be back on the home platform — don't linger
  around the far side. At most **6 seconds** of quiet time away from home. Then resume
  the patrol sweep.

The preview prints `[preview]` lines reporting these duties (falls, ghost chases,
unconfronted visitors, overstayed absences, patrol coverage) so you can watch a run
and see which duty broke.

## State fields

| field | type | description |
|-------|------|-------------|
| `self_pos` | Vector2 | guard center |
| `velocity` | Vector2 | current velocity (y-down positive) |
| `is_on_floor` | bool | standing on ground this frame |
| `platforms` | Array[Rect2] | all platform rects (top surface = rect.position.y) |
| `walls` | Array[Rect2] | tower rects standing on platforms (`[]` when none) |
| `home_rect` | Rect2 | your home platform (patrol + return target) |
| `intruders` | Array | `[{id: int, pos: Vector2}, ...]` — the full roster, always |
| `vision_range` | float | watch radius (varies by yard) |
| `world` | Node2D | physics space with the colliders — you may raycast against it |
| `dt` | float | timestep (1/60 s) |
| `t` | float | elapsed time (s) |

Note `intruders` always lists every visitor with its true position — deciding which of
them you can *see* is your job, not the state's.

## Physics constants

- `SPEED = 200` — horizontal speed at `move=1`
- `JUMP_VELOCITY = -400` — jump impulse (upward)
- `GRAVITY = 980` — y acceleration per second squared
- Guard capsule: radius 12 (center is 12 above the ground when standing)
- Watch length: 1800 frames (30 s at 60 Hz)

## What is fixed

Your deliverable is the controller under `res://logic/` — build it against the state
interface above. The rest of the project is the game itself: your AI has to work with
it exactly as it stands here. You can change anything locally while debugging, but
changes outside `res://logic/` are debugging aids, not part of your deliverable.
\end{promptbox}

\subsection{Repo task}

This example shows a Repo task implemented within an existing project.

\begin{promptbox}{Task Brief --- AMSG Character Movement (Repo)}
# AMSG — the character movement component

You are working inside **AMSG (Advanced Movement System Godot)**, a Godot 4 third-person
character kit (MIT; upstream notes in `README_UPSTREAM.md`). The character rig is the real thing —
a skeletal mesh with its animation system, a camera component, a stair sensor shape, collision
shapes, mantling and pose-warping addons — and all of it is driven by the state of ONE component
sitting on the rig.

Right now that component is **missing its behaviour**. The rig loads, the world runs, but the
character does not move, does not fall, does not crouch, does not jump. Your job is to rebuild the
movement engine behind the frozen interface.

## What you deliver

A single file:

- **`res://addons/AMSG/Components/CharacterMovementComponent.gd`** — the movement component. Its
  method bodies have been removed; the class, every export, every state variable the rest of the
  rig reads, and every method signature (with default arguments) are declared in the stub. **Keep
  the interface exactly as declared** — the animation blender, the camera component, the player
  controller and the gameplay components all read this component's state and call it through
  these declarations.

You may add helper scripts under `res://addons/AMSG/Components/` and `preload()` them, but the
file above is the deliverable.

## How it connects (interface facts)

- The rig scene (`res://AMSG_Examples/Character/mixamo_character.tscn`) wires the component's ten
  reference slots to its own nodes and assigns the six movement-data resources. The evaluation
  world rewires all six data slots to its own distinct `movement_values` resources and sets the
  world's `deacceleration` — exactly as `test_arena/preview.tscn` demonstrates.
- The caller (see `AMSG_Examples/Player/PlayerController.gd`) drives movement by calling
  `add_movement_input(direction, speed, acceleration)` on every simulated step **while movement
  input is held**, picking the speed and acceleration for the current `gait` tier out of your
  `current_movement_data`. When input is released the caller simply **stops calling** — read the
  `add_movement_input` doc comment in the stub for what idle means.
- `rotation_mode`, `gait` and `stance` are plain properties that the caller and other components
  **write from outside at any moment while the character is moving** — including components of the
  frozen rig writing them back (see `CameraComponent.gd`). `jump()` is called when the player
  jumps.
- The animation blender (`addons/AMSG/Components/AnimationComponents/AnimationBlend.gd`) reads
  your component's state every simulated step to drive the animation parameters — what it reads
  is visible in its code.
- How the component is supposed to behave is in the surrounding code and the stub's own doc
  comments — read the rig scene, the data resources, the consumers. **The behaviour contract is
  not spelled out here on purpose; reconstruct it from the codebase.**

## What the character must do (functional expectations)

- Move in three gaits — walking, running, sprinting — at the speeds of the **active movement
  data**, and switch data immediately when `rotation_mode` / `gait` / `stance` change, no matter
  when or from where they are flipped.
- Coast to a stop when movement input is released.
- Crouch and stand back up; **when something overhead blocks standing up, the character must not
  rise into it, and must not gain height from a jump either — full height comes back once the
  blockage is cleared.**
- Step up stairs and ledges no taller than the configured maximum stair height (see the stub's
  exports), from whatever direction the character happens to travel.
- Jump only when on the ground.
- After ground contact is lost there is a **0.1-second confirmation window: until the fall is
  confirmed the character is not treated as airborne and takes no gravity** (this is what makes
  stepping down a small ledge feel solid).

## The world varies

The evaluation worlds are built procedurally — floor layout, steps, gaps, overhead geometry and
the input script (which keys are held when, which modes get flipped where) vary from run to run.
Your component has to hold up under whatever legal driving pattern the world presents.

## Trying your work

```
godot --headless --fixed-fps 60 --path . res://test_arena/preview.tscn -- --seed 1
godot --headless --fixed-fps 60 --path . res://test_arena/preview.tscn -- --seed 7
```

The preview builds the public baseline arena, drives your component through the real caller
pattern and prints `[preview]` readouts (measured speeds next to the commanded table values, the
step-up height, the jump rise). Run it windowed (F5) for a follow camera. `test_arena/` is a
debugging aid the project ships for you — it is not part of your deliverable. The upstream demo
map (`AMSG_Examples/Maps/MovementTestMap.tscn`) is also intact if you want to drive the character
by hand in the editor.

## Where your work ends

Your deliverable is exactly `res://addons/AMSG/Components/CharacterMovementComponent.gd` (plus any
helpers you add under `res://addons/AMSG/Components/`). Everything else — the rig scene, the
animation blender, the camera, the player controller, the addons, the data resources, the project
configuration — is the game itself; your component has to work with it exactly as it stands here.
While developing you may change anything locally (add prints, tweak the arena, try another seed),
but changes outside your deliverable are debugging aids, not part of your deliverable.
\end{promptbox}

\section{Construction and Review Standards}
\label{app:prompts}

The pipeline in Section~\ref{subsec:construction} uses four construction agents and two human-curated stages. The following prompts are given to the prototyping, blueprint, implementation, and review agents. The final two documents define the screening and human-review standards.

Placeholders in braces are filled per task. In the construction prompts, an \emph{axis} is one named capability a scenario is built to press, a \emph{rung} is one calibrated depth on such an axis, and a failure is \emph{charged} to the axis whose contract broke first. This vocabulary is specific to task construction and differs from the capability classes in Section~\ref{subsec:profile}.

\subsection{Construction-agent instructions}
\label{app:prompts:construction}

The four roles correspond to the pipeline stages of Sections~\ref{subsec:construction} and~\ref{subsec:gate}. Two standards govern them. The prototyping, blueprint, and implementation agents work from the authoring standard, which is where the stage-by-stage procedure lives. The review agent receives only the acceptance standard, and that separation is what keeps its perspective independent of how a task was built.

\subsubsection{Prototyping agent}
\label{app:prompts:prototyping}
\begin{promptbox}{Prototyping Agent}
# Prototyping agent - instructions

## 0. Role and boundary

You work before task construction proper begins. Your input is a candidate mechanic specification (a description of the mechanic plus an initial tier assignment: atom, combo, or repo). Your output is a feasibility report. There is only one question to answer: can this mechanic carry a deterministic criterion, and are the readings of a proper implementation and a naive implementation separated by a stable margin?

You do not write the brief, you do not build a library task, and you do not produce the formal calibration artifacts. What you build is a throwaway probe, archived once used and never admitted to the library.

## 1. Measure engine behavior, do not assume it

Every engine capability the mechanic depends on - physics queries in headless mode, signal ordering, navigation-service behavior, coroutine settlement order - is measured in a minimal project before anything else proceeds. "The documentation says it should work" is not evidence.

The same measurement confirms that these capabilities reproduce deterministically in headless mode: the same input, run three times, yields bit-identical readings. If a capability injects nondeterminism into the tested behavior itself - for instance, a navigation map whose rebuild runs on a worker thread, so the same probe observes different frame delays - find the switch that pins it down. Drop mechanics that cannot be pinned.

## 2. Two implementations and the separation margin

The probe compares the readings of two implementations over the same set of parameters.

- The proper implementation uses the abstractions the engine provides, reads current runtime state every frame, and sequences the mechanic correctly.
- The naive implementation is the strongest version a red-team reading would write: the classic bug patterns - planning once, counting down timers in parallel, counting frames instead of using the time base, ordering once at the top of a round, writing derived quantities straight back into base quantities. It must not be carelessly broken code. It has to be the version an experienced engineer would naturally write without understanding this mechanic.

Both implementations run across the parameter grid. A candidate proceeds only when the readings separate with a stable margin: the proper implementation passes everywhere without sitting on the tolerance line, and the naive implementation fails on the intended axis. Any of the following is a negative result and is reported as such:

- the readings overlap, or the separation holds only at isolated parameter points;
- the naive implementation also passes, so the mechanic is not a hard gate;
- the proper implementation cannot pass, so either the criterion is too strict or the mechanic cannot be implemented stably on this engine.

## 3. Rechecking the form of the criterion

Human screening has already judged whether this mechanic can carry a deterministic criterion. Your job is to recheck that judgment and to write out what this task's hard-gate assertion actually looks like.

The criterion must be mechanism-form: it asserts a unique correct value, a unique legal ordering, a conserved quantity, or an exact rollback.

Asserting that some bound holds - no more than N leaks, survival, coverage above a threshold, a budget not exceeded - is threshold-form. Its failures cannot be attributed to a specific contract, so it may serve only as an admission gate or a saturation gradient, never as a hard-gate assertion that carries discriminative weight.

A candidate for which no mechanism-form assertion can be written does not go to the blueprint stage.

## 4. Rechecking mechanism ownership

The implementation of the tested mechanic - the mechanism mathematics, meaning the integrator, the scheduler, the ledger, the solver, the state machine - must live in the code the agent delivers, rather than the agent merely scoring preferences over actions the game already offers. The test is not whether the output happens to be a choice. It is whether producing the correct output requires mechanism state on the agent's side.

The discriminator is a random stub: replace the agent-side code with a well-formed random stub, run the criterion, and see where the failure signature lands.

- If only outcome bounds break (leaks, win or loss, casualties), the mechanism in fact lives on the game or judge side and the agent is merely a player. Do not build the task.
- If mechanism contracts break (conservation, timing, ordering), the implementation of the mechanism is in the agent's hands and the candidate qualifies.

The shorthand: delete the agent's code, and ask whether the game gets dumber (it plays badly but nothing is missing) or crippled (something stops happening altogether). Only the latter can become a task.

This pass with the random stub is an early, cheap version of the mechanism-ownership recheck that happens before admission. It does not replace that one: the check is run again after the task is built, because the pressures of building push mechanism content from the agent's side toward the judge's side.

Confirm also that the mechanic can be stated in the vocabulary of a game-development role ("implement flocking avoidance", "implement an input buffer"), rather than "play some game well".

The level the mechanic belongs to is rechecked here as well. Three vocabularies: engine subsystems (physics, navigation, rendering, input); genre-general gameplay mechanics that have names and transfer across games (steering and avoidance, input buffering, production queues, line of sight, turn order, status and modifier settlement); and one specific game's content systems (its data tables and rule instances). The tested mechanic must belong to the first two. The third may serve only as a source of difficulty, never as the channel through which score is read: when difficulty lives in the content system and score is read off wins and losses, depth of understanding and quality of strategy cannot be separated within the score.

## 5. Mechanism thickness and the sources of difficulty

The mechanism surface has to have thickness, favoring cross-frame state machines, temporal scheduling, and networks of conserved quantities. A thin arithmetic surface (a few clamps and sums) is accepted only when it serves a structural purpose, and it must be flagged as a difficulty alert in the report. Human screening has already made a first assessment of thickness; you recheck it.

Reducing the mechanism surface on the grounds that it would be easier to implement or easier to calibrate is forbidden. Difficulty has exactly three positive sources: mechanism thickness, the awkwardness of the legal call patterns, and the need to read an existing real project.

Depending on the engine is not the same as being difficult. Engine dependence is pursued for positioning and ecological validity. Expectations are never raised on the grounds that "it uses the engine, so it must be harder".

## 6. Verdicts and exits

- Stable margin, mechanism-form criterion, mechanism owned by the agent: the candidate proceeds to the blueprint stage and the report becomes blueprint input.
- Unstable margin: change the parameter band or strengthen the naive implementation and try once more. If it is still unstable, return the candidate to the pool with the failure mechanism written down.
- Mechanism not owned by the agent, or only a threshold-form criterion can be written: the candidate is not built, with the reason recorded.

## 7. Report template

1. Candidate name and initial tier assignment.
2. Engine-behavior findings: what was measured, how, whether determinism holds, which switches have to be pinned.
3. Probe construction: how the core of the mechanic was isolated, and which dimensions the parameter grid spans.
4. The two-implementation reading matrix (proper and naive against each parameter point: pass or fail, and which axis a failure lands on).
5. The separation margin: how large it is, and at which parameter points it is thinnest.
6. Recheck findings for criterion form and mechanism ownership, including the random-stub signature and the level judgment.
7. An assessment of mechanism thickness, with a difficulty alert where warranted.
8. Recommendations and known risks for the blueprint stage.
9. Verdict: proceed to blueprint, try once more, or do not build.

Report conclusions and the key readings. Do not paste long stretches of code.
\end{promptbox}

\subsubsection{Blueprint agent}
\label{app:prompts:blueprint}
\begin{promptbox}{Blueprint Agent}
# Blueprint agent - instructions

## 0. Role and boundary

Your input is the candidate mechanic specification and the feasibility report. Your output is a blueprint that can be built from. The blueprint is the sole design authority for the implementation stage, so it has to settle all of what this task tests, how it is judged, how the scenarios are laid out, and where difficulty comes from, and it has to carry the measured evidence from two machine checks.

You do not build the library task, you do not write the final brief, and you do not produce the formal calibration artifacts. But the key judgments in the blueprint must be backed by evidence from actual runs: the evidence for both machine checks in Section 8 is produced at this stage and may not be deferred to implementation or review. Both failure modes they catch - a criterion anchor that can be translated straight out of the brief, and a difficulty axis that turns out to be empty - only become visible after the task is built, and neither can be repaired then. The cost is a whole round of scenario work thrown away. The implementations used for these checks are throwaway probes. They are not the same artifacts as the formal calibration set built during implementation, they never enter the library, and they never enter the calibration record. Every new repo-tier and combo-tier blueprint must contain both checks. For an atom blueprint, mechanism complexity decides whether they are needed.

## 1. Deliverable

One blueprint, containing:

1. the definition of the tested mechanic and its capability name (in the vocabulary of a game-development role, never naming a game);
2. the form and boundary of the deliverable (which files the agent owns, how the game calls them);
3. the layer-by-layer split of the three-layer information boundary (Section 2);
4. the criterion design: the form and the anchor of the hard-gate assertion, plus the answer to the anchor question (Section 4);
5. the difficulty recipes (at least two heterogeneous ones, Section 5);
6. the scenario design (the list of structures, which axis each scenario arms, the safe numeric bands the seeds perturb, Section 6);
7. the measured evidence from both machine checks (Section 8);
8. the tier assignment and the argument that the form of the task complies (Section 7);
9. known risks and the points the implementation stage has to calibrate with particular care.

## 2. The three-layer information boundary

Information given to the agent is split into three strict layers, and the blueprint has to assign every fact to one of them.

Layer 1, functional requirements and fixed rules - what the game needs, the fixed rules of the running game, and the fact that the arena is generated procedurally on every run. This layer must be disclosed.

Layer 2, situation difficulty and concrete design - how hard a situation is, whether anything changes mid-run, how the scored scenarios escalate. This layer is not disclosed.

Layer 3, the evaluation mechanism - how judging itself works. This layer is never mentioned.

The dividing line: generation on every run belongs to layer 1, while mid-run dynamics and the direction of escalation belong to layer 2. Two reasons.

"The arena differs every run" is a fixed rule of the game itself. A brief that stays silent about it only manufactures measurement noise over whether the agent bothered to read the project files, and it lets failures hide behind the defense that the work order never asked for generality, so a minimal solution was the rational move. Once it is disclosed, the claim strengthens to: generation on every run was stated plainly, and the solution still broke in harder or dynamic situations.

What the leakage evidence actually leaked was "the terrain changes, the door closes" - mid-run dynamics - which amounts to handing over the architectural answer of replanning every frame. Disclosing generation on every run does not touch that evidence.

Where the difficulty gap comes from. The example scenario is a standard, easy situation; the agent develops and self-tests against it. It knows the arena is generated every run, but not that harder situations or mid-run dynamics are coming. The scored scenarios are harder situations under the same functional requirement, applied only at judging time. A naive implementation - planning once, improvising - therefore passes on the example scenario and breaks on the scored ones, while a proper implementation that reads current runtime state every frame survives without ever being told.

Fairness constraints. Violating any of these makes the task a harness defect rather than a capability test.

1. The state interface is identical on the example scenario and the scored scenarios. Every information channel the scored scenarios rely on is offered on the example scenario too. The interface does not change; it merely leaves caching-once exposed.
2. The proper solution must pass without being told. It relies only on reading current runtime state to choose the next step, never on advance knowledge that something will change.
3. The scored scenarios may only add harder situations under the same functional requirement, never a new functional requirement. Corridors may narrow, targets may multiply, a mechanism may fire mid-run. What they may not do is suddenly require a capability absent from the example scenario.
4. Systems are disclosed; instances are not. A scored scenario may contain content or a mechanism the preview never demonstrated, provided it belongs to a rule system already disclosed in layer 1 of the brief and is observable through exactly the same state channel as on the example scenario. What is disclosed is the rule system, not the instance. Introducing an undisclosed system - semantics absent from the state contract - amounts to demanding mind reading and is forbidden. Two honesty constraints follow. Every rule written in layer 1 must genuinely occur among the scenarios this task actually runs, so no decoy rules that never appear. And every violation the criterion punishes must be traceable to a rule that was either disclosed or discoverable in the agent-visible project; if it traces to neither, it has fallen into layer 3 and become a scoring basis only the task author knows.

## 3. Brief narrative and instruction register

Instructions stay open-ended. The brief describes the behavior to achieve. It does not name specific abstractions, nor APIs, properties, or function names. The agent chooses its own abstractions, because the criterion looks only at behavioral consequences and takes on no attribution debt for "was this abstraction used". For the same reason, the tested capability is never claimed to be the use or non-use of a particular engine interface, which is an attribution dead end: the steady-state usage of an engine subsystem has to be disclosed in layer 1 anyway, and once disclosed it becomes a one-step consequence of a rule sentence.

The narrative is a work order. Briefs are always framed as implementing a system or a feature for the game - an automatic damage-control system, a labor-assignment system, a combat decision module, all functional modules of the game itself. Never "write an AI or a bot to play this game." For the same deliverable, the first framing is an engineering work order and the second is a game-playing bot. This holds across all three tiers.

## 4. The criterion-anchor rule (the blueprint must answer this)

The form of the criterion was settled at the probe stage. What this section settles is the anchor. Among assertions that are all mechanism-form, discriminative power depends on where the correct behavior is derived from.

- Discriminative: the criterion is anchored to a temporal snapshot or to state-inheritance semantics over runtime observables. The correct answer is reached by applying an already disclosed family of static rules to a dynamic situation that spans frames, stages, or phases. The brief can disclose the rule family (layer 1) but cannot spell out the correct output for each case, because the dynamic situation itself is undisclosed layer-2 information.
- Bound to be translated: the correct behavior the criterion anchors to is a one-step consequence of a rule sentence in the brief. Fair disclosure and the solution specification are then the same object, and the task is expected to saturate.

The question to answer: can the correct answer the judge's assertions anchor to be obtained by translating, in one step, the rules the brief already discloses?

If yes, there are three options: move the anchor into a dynamic situation, admit the task honestly as coverage (carrying no discriminative weight, with a difficulty alert in the calibration record), or do not build it.

Tightening the wording cannot fix a misplaced anchor. Both repair paths have been tried and neither worked: removing walkthroughs and hints while keeping the rules themselves, and withdrawing the reference unit tests while deleting the paragraph of the brief that translated straight into the answer. To judge whether tightening will help, first ask whether the anchor can move. An anchor in a dynamic situation responds to tightening. An anchor on information that has no definition left in the project once the system is carved out, and therefore has to go into the brief, is structurally unsolvable, and no scenario work should be spent on it.

A note on genres. Genres that advance in real time naturally offer two separable dimensions: the algorithmic content can be disclosed in full, while the correct temporal mounting exists only inside the frozen runtime. In genres that settle in discrete turns, "timing" is phase order, and phase order is itself the tested content, so the two dimensions collapse onto the same anchor and no change of carve-out surface will separate them. Before committing to a turn-based real-project task, check whether the anchor can be decoupled at all.

## 5. Two-recipe discipline

A criterion that is too loose degenerates the task into a binary gate: use a certain abstraction and score everything, or don't and score nothing. Gradients come from perturbation axes - tightening resource budgets, making the scored situations harder, scaling up, applying pressure mid-run - so that a gap in pass and fail also opens between a mediocre implementation and a proper one.

The difficulty of every producer-side task draws on at least two heterogeneous recipes.

- Engine time base and scheduling semantics: under a stretched time step, time scaling, or pause semantics, implementations that count frames drift.
- No fixed doctrine: the same policy produces opposite correct behavior in two coupled situations, and there is no way to know which one you are in.
- Necessary but not sufficient, in two layers: choosing the right system is only the necessary half; it still has to be driven correctly.
- Undisclosed but legal call patterns: same-frame concurrency, re-entry, extreme values, call ordering.

A single recipe - especially the time-base axis alone - collapses the gradient back into a binary gate. Whether two recipes genuinely hold is machine-checked by the catch rate of the mutant family: if a task claims to press the time-base axis but a degenerate implementation that simply counts frames passes, that axis is fictional.

## 6. Scenario design

The anti-pattern is building the scored scenarios as one branch plus narrow-band jitter over seeds. N seeds are then N slightly repositioned copies of the same trap, carrying about one bit of information; getting one mechanic right clears them all, which is structurally a binary gate.

The rule in force: the structure, topology, mechanisms, and events of a scenario are designed by hand, and seeds perturb numeric values only within bands that never break solvability. The expensive part, structural legality, is done once by hand; the cheap part, the numeric band, is left to seeds.

- Every task keeps exactly one example scenario. What the preview in the agent-visible project runs is its twinned copy.
- Every scored scenario declares explicitly which axis it arms and at which rung. The scenario name is only a display label; machine identity lives in the axis and rung declaration. This is mandatory for every scored scenario of a combo task. An atom task's judge does not consume the axis declaration and is exempt.
- Rung names must be traceable. A rung on an axis has to correspond to a scored scenario that genuinely exists in that axis's origin task, so the rung name declares where the triggering condition of the defect it is meant to catch was transplanted from. Recalibrating numeric bands and geometry for the new situation is normal, and the difference is recorded in the calibration record. When a construction has no counterpart in the origin task, either backfill it there as a new scenario or change this task to match an existing one. Inventing a rung the origin axis does not have is not allowed.
- The attribution contract: on a single-axis scenario, the axis charged with the failure must equal the armed axis; on a coupled scenario this relaxes to the charged axis belonging to the armed set. A failure charged outside the armed set is a construction error. A coupled scenario must also carry a scoring signal that the single-axis scenarios do not already provide, and the argument for that incremental signal goes into the calibration record.
- Coupled scenarios need two probes. A tailored probe that is correct on axis A and naive on axis B must be charged to B, and vice versa. A coupled scenario holds only when each axis can be attributed independently.
- Selection discipline. Structural-depth axes (navigation, flocking avoidance, ballistics, where solutions have a real ladder of rungs) are laid out rung by rung. State-reading axes (cooldown, hitstun, death, selection, where "read the state every frame and act on it" is already the ceiling) have one rung, which is also the last. Coupling is attempted only for combinations of a structural-depth axis with a state-reading axis for which there is an explicit behavioral hypothesis: the structural axis forces a multi-frame committed maneuver, and the question is whether another discipline lapses during it. Never take the Cartesian product of axes and rungs.
- All-axes-armed scenarios. Once the single-axis matrix is filled out, one scenario may be stacked on top in which every axis takes its deepest calibrated rung. It is not an attribution instrument; what it reads is degradation under load. Hence at most one per task, added only when that task falls short on discriminative weight.
- Defenses against unsolvable instances: structural legality is verified by hand once; numeric bands are chosen in ranges that never touch connectivity or solvability; calibration curation is the backstop. Nothing is random at evaluation time.

## 7. Tier and form discipline

The three tiers.

- Atom: the minimal task for a single mechanic, the anchor for capability calibration.
- Combo: several coupled mechanics implemented in a minimal purpose-built game, with no need to read an existing real project. The mechanics may be transplanted from atom tasks or original.
- Repo: integration into a real project. The need to read the real repository is a defining property, not a residue - it is what carries ecological validity. The information source for the behavioral contract is the repository itself. A "real-project task" whose contract is written out in the brief is not merely on the easy side; it has left the tier definition and become a combo task with a real skin, and review rejects it on eligibility grounds.

When a combo task and a repo task are paired, only one variable may differ: the need to read the project, and code volume. The attribution instruments are within-tier axis attribution and the difference in readings for the same axis between its atom setting and its combo setting.

A prohibited form: modules that are pure functions with the interface contract fully given are no longer built. The mechanism is that in this form, fair layer-1 disclosure and the solution specification are the same object. Fairness requires that the rules being enforced be disclosed, and once the rules are written out in full, the behavior of a pure function is fully determined for every input; the scored scenarios only change the input, and the required behavior is already on paper. Tightening the wording does not help, and neither does awkward composition (re-entry, concurrent call ordering).

The callee delivery form remains legal, but the contract has to satisfy one of two conditions: it is buried in a real project so it has to be read, or the mechanism is thick enough to constitute an implementation wall on its own (interacting couplings, several simultaneous constraints, rather than transcription from a table). Both still have to clear the criterion-anchor check.

The three ways a real-project task can be formed all start from something missing, never from something broken.

- Complete a partial system: the repository genuinely lacks or has not finished the system, and it is completed to specification. The proper solution is written by the author.
- Carve out and reimplement: upstream has a complete implementation, the task version carves the whole subsystem out, and the agent rewrites it against the specification. The proper solution is the original upstream implementation, and the fact that it passes the criterion is direct evidence that the criterion does not kill correct solutions.
- Extend within the coupling: one mechanism is added inside the constraint network of the existing systems without breaking any conservation that already holds. The proper solution is written by the author.

Multi-system coupling is a hard gate for tasks that carry discriminative weight: the deliverable of a real-project task must span at least two heterogeneous system axes. The axis-type vocabulary: spatial behavior (navigation, avoidance, pursuit, interception); temporal scheduling (attack clocks, cooldowns, turn scheduling); resource ledgers and pool arbitration (conservation ledgers, a shared scarce pool from which at least two systems - each with its own state machine and failure mode - draw, where calibration must also demonstrate that a rung exists on which any fixed doctrine fails, or else the pool is only scenery); state machines and mode switching (re-evaluating engagement, promotion and demotion, commitment and withdrawal); conservation and network propagation (supply graphs, spreading, conduction). "Heterogeneous" means the two axes belong to different types; two rules within one type do not count. A narrow slice of a single system carries no discriminative weight and may serve only as a saturation gradient, which the calibration record has to state.

Coupling is not a bot. What is carved out is the mechanism itself on the engine side, the criterion is anchored to the mechanism contract of each axis, and play outcomes serve only as an admission gate.

The brief for a real-project task is a minimal interface specification, and writes only four things: where the deliverable sits (which system is missing, which files); the interface facts (how the game calls the deliverable - constructor and static method signatures, signal and advancement conventions, the direction of data flow); the layer-1 disclosure; and the fallback facts (constants and rules the criterion depends on that are not discoverable inside the project, each as the simplest possible factual sentence).

Any behavioral contract that is discoverable inside the project stays out of the brief. The information source for the contract is the repository itself: caller code, constant files, neighboring systems that were not carved out, data tables. Writing the contract into the brief short-circuits the need to read the project into "read the spec and transcribe", which dismantles one of the three sources of difficulty on the spot. Skeleton and stub comments are held to the same standard as the brief: a lean brief whose stub comments still spell out the contract changes nothing.

How this differs from symptom-driven defect repair: those tasks describe a symptom (something is broken), whereas this form gives an interface plus a repository (a system has to exist, and what it looks like is for you to read). Never write the algorithm or the order of implementation, or the task degenerates into transcription. The behavioral contract stays out of the brief as a whole - when it is called, which coroutine settles first, which state is mutable at that moment, along with how long a beat lasts, how conservation is accounted for, and how state transitions - all of it has to be read from the repository code.

There is no gate of the form "an independent implementation must pass the criterion", because that would reward writing the specification in more detail and choosing thinner mechanisms. There is no blind-writing step either (no uninformed implementer writing a second proper solution to validate the specification): failure is the default outcome for this class of task and carries no information, and insisting on it would leave tasks without a proper solution. Its two functions are carried by means that spend no model time - informational closure by the discoverability check, criterion neutrality by the catch rate of the mutant family.

## 8. The two machine checks (evidence produced at this stage)

Check 1, criterion-anchor discoverability. List every fact the judge's assertions depend on (constants, rules, temporal semantics) and dispose of each under one of three classes.

- Class A, discoverable in project code: never written in the brief; the blueprint gives independent evidence (where the fact lands, in which file). Skeleton and stub comments do not count as evidence, since they belong to the brief.
- Class B, currently knowable only from the brief: move it back into the project where possible (keep a neighboring upstream implementation, a data table, a constant). If it cannot be moved back, it is demoted to class C.
- Class C, present nowhere (constants and rules that vanished along with the carve-out): it must go into the brief as the simplest possible factual sentence, as a fallback.

No scoring basis that only the task author knows may be built on. Fix the specification or change the assertion; this line admits no exceptions. For each contract family, also answer: can the agent infer what has to be handled, and if it cannot, on what basis does a scored scenario charge it?

Check 2, the empty-axis check. Every difficulty axis gets one single-defect naive implementation, run across all scenarios as a measured matrix (implementation against scenario: pass or fail, plus the axis charged). The axis has to break, with zero collateral damage on other axes. This is measured, not argued. These single-defect implementations are throwaway probes for this stage, not the formal mutants of the implementation stage. When an axis is suspected of being empty, there is a cheap discriminator: run two frozen configurations that differ only in that constant on the same situation and compare the result fingerprints.

## 9. Verdicts and exits

- Both checks pass, the anchor sits in a dynamic situation, two recipes hold, and the form complies: the blueprint goes to the implementation stage.
- The anchor can be translated in one step: move the anchor, or admit the task honestly as coverage with a difficulty alert, or do not build it.
- An axis is empty: change the scenario assignment or the numeric band and rerun the check. If it cannot be saved, drop that axis and recheck whether two recipes still hold.
- A complete implementation follows directly from the brief without requiring any reasoning about runtime interactions: drop the candidate.

## 10. Report template

The blueprint is itself the deliverable. When reporting, give: the mechanic and its capability name; the layer split findings; the answer to the anchor question; the measured matrices and conclusions from both machine checks; the difficulty recipes and the structure of the scenario set; the tier assignment and form-compliance finding; known risks and the calibration priorities left to the implementation stage; and the verdict (go to implementation, re-anchor, or do not build). Report conclusions and the key readings. Do not paste long stretches of code.
\end{promptbox}

\subsubsection{Implementation agent}
\label{app:prompts:implementation}
\begin{promptbox}{Implementation Agent}
# Implementation agent - instructions

## 0. Role and boundary

Your input is the blueprint and the feasibility report. You turn the blueprint into a task that can enter the library: the judge side, the agent-visible side, the four kinds of calibration artifact, and a calibration record at the granularity of a single test case.

What you hand over has to survive an uninformed review agent rerunning it through the regular container pipeline. That agent does not take any reading you report on trust, so every number in the calibration record has to be one you actually produced.

## 1. Deliverables

1. The judge side: the authoritative level definition (the single source of truth for geometry and state), the assertions, the driver, and the determinism settings.
2. The agent-visible side: a playable game project containing the preview and debugging interface, the stub for the deliverable, and the brief.
3. The brief: the game's functional requirements and fixed rules, plus the boundary of the deliverable, and nothing else.
4. The four kinds of calibration artifact: the proper solution, the naive solution, the family of single-capability mutants, and the behavior-preserving control.
5. The calibration record: for every artifact, every scenario, and every seed, pass or fail; the margin figures for the proper solution; and which axis each failure signature lands on.
6. The recording layer used for presentation (which takes no part in judging, see Section 8).

## 2. Information discipline (hard constraints)

### 2.1 The leakage red line

Everything the evaluated agent can read - the brief, every comment in the agent-visible project, the comments in the stub - describes only the game's functional requirements and its layer-1 fixed rules, and never mentions the evaluation mechanism.

- Never say "there are other arenas on the evaluation side", "the seed will be changed and rejudged", or "you will be tested on a different scenario". Those are layer-3 evaluation strategy.
- Never say "you will be scored" or "you will be graded on a different layout".
- Never give architectural instructions such as "don't hardcode", "write general code", or "requery every frame instead of caching". That hands over the answer. The dividing line runs here: "it has to work on any arena the game generates" is a statement of requirement (it says what is wanted), belongs to layer 1, and is legitimate. The same fact translated into how to write the code is forbidden.
- Do say what the game needs: the enemy has to reach the goal past a level with a door; the boss needs aggro, approach, an attack cooldown, a hit reaction, a death trigger.
- Do state that the arena is generated procedurally on every run (layer 1): the game builds the arena procedurally, and the walls, the doorway, the start and the goal are laid out differently from one play to the next, with the preview wired to one example. State the fact about the game itself, and stop there - no "so make it general", and no mention of testing.
- The narrative is always "implement a system or feature for the game", never "write an AI or a bot to play this game". Behavior is described by the effect to achieve, without naming specific abstractions, APIs, properties, or function names.

Why. The intent to generalize is established by the work order: a real work order does say "the arena is generated every run", and generalizing is what the engineer is there for. That is a requirement, not a hint. Whether the implementation of that generality survives harder or dynamic situations is the capability under test. Telling the agent "there are other tests, don't hardcode, the arena changes mid-run" is handing over the answer, and what leaks is precisely the layer-2 dynamic that the door closes. A real work order also would not say "we will test you on a different set of maps."

### 2.2 Three audits (all of them)

Audit 1, wording. Sweep the whole agent-visible project and the brief, and rewrite any hit in neutral terms. There are four classes of word to sweep for: words naming scoring or a scorer; words naming "other situations on the evaluation side"; words naming how we measure (judging, recording, tier, scenario, and similar internal vocabulary); and architectural instructions demanding robust or general code or forbidding hardcoding.

A variable name that holds a seed is harmless engineering vocabulary and may stay. Descriptions of evaluation strategy such as "judged across seeds" or "the seed will be swapped" have to go. The agent-visible twin of the level definition carries no tier or scenario parameter: the selection of scored scenarios lives only on the judge side, while the example scenario the preview runs is itself visible to the agent. A comment in the twin saying "differs from run to run" is a layer-1 fact, legitimate, and should agree with what the brief discloses.

Audit 2, structure. Compare the key sets of the two level definitions. The agent-visible twin must not contain a key that only the judge side consumes, because the key name itself leaks the concept. In one earlier case the twin still held a boolean key meaning "the door closes" together with a trigger coordinate, which gave away the layer-2 dynamic. Judge-side keys are read through a default at the point of use, so the twin does not break when a key is absent. This is different from a tolerance constant needed for fidelity: the latter is a known trade-off the preview requires, the former is pure judge residue and costs nothing to delete.

Audit 3, comments. Visual logic the preview does not use, together with its comments, goes into the recording layer rather than the agent-visible project. Comments the agent can see may only describe the behavior of the preview itself. In one earlier case a comment in the visual code wrote out "the door collider the judge adds mid-run", which gave away the layer-2 dynamic inside a file the agent can read.

### 2.3 Never feed a false completion criterion

The shared solve instruction is a task-neutral operational guide used across the library, and the completion criterion always points back to the functional requirements in the brief. The preview is positioned only as a debugging aid (it reports what happened, including any rule violation). Never write "iterate until the preview passes": that actively feeds a false criterion in which passing the example scenario equals finishing the job, contradicts the requirements in the brief, and invites a minimal solution that treats a single preview run as the finish line. The layer-1 disclosure lives in the brief, and the shared instruction neither repeats nor adds to it. Changing the shared instruction means changing the solve input, which invalidates every existing workspace and forces a rerun.

### 2.4 Constrain the deliverable, not file permissions

Never write "read-only" or "do not edit". Protection against tampering has always come from the order in which files are overlaid at judging time, never from wording about permissions. Wording about permissions carries three real costs. An agent that wants to test itself on more than one instance - changing the preview seed, reshaping the arena to stress its own code - has no legitimate channel, so the tested behavior of whether it builds its own test setup cannot be expressed. The brief has already disclosed that the arena is generated every run, yet no means is provided to preview a second arena, so the disclosure and the tooling contradict each other. And the environment would then never have offered any way to try more than one arena.

What is fixed, in the brief, is always written as the boundary of the deliverable. The deliverable is the implementation at the named paths. Everything else is the game itself, and the agent's code has to work with it exactly as it stands. Local changes are free, and changes outside the delivery paths are debugging aids that are not part of the deliverable. Stub comments say "build your implementation on top of this; it is not itself your deliverable." The second half - that the integration target is fixed - has to be present: deleting "read-only" without establishing the boundary of the deliverable leaves the task ambiguous (widening the doorway would also count as finishing), and having a reasonable reading score zero is a harness defect rather than a capability test.

## 3. Criterion implementation discipline

Fully black box. The criterion observes only observables of runnable behavior (positions, health, targets, attack timing, geometric clipping). No source inspection, no language model in the scoring path. Assertions are parameterized and can be rejudged under a different seed. The scored scenarios and their seeds never enter the solve environment; they are injected by the evaluation side only at judging time.

Read the module through the game (this rule does not bend). The criterion drives a complete game scenario for a full run, and the assertions read only runtime observables and the event stream (who struck first, on which frame something expired, conserved quantities, release ordering). Never make the return value of the delivered module the object being scored, and never call it in isolation as a unit test. Breaking this degenerates the task into property-based module testing and dismantles what distinguishes this benchmark on the spot.

Two families of assertion. Judging keeps the library's native register: every test case yields a binary pass or fail, plus one axis charged with the failure.

1. The contract of the new system: the behavior the specification requires holds on every frame under the undisclosed but legal call patterns of the scored scenarios - concurrent application in the same frame, re-entry, extreme values, a stretched time base. The example scenario issues only gentle calls. The semantics of a scored scenario is a more awkward legal call, not a stronger opponent.
2. No regression in existing behavior: driven by the same script, the conserved quantities and key invariants that the original upstream version produced still hold after the agent's changes. The reference is the batch of numbers produced during construction by running the original upstream code on the same scenario and storing them. It does not rely on tests shipped with the repository. A regression violation is one of the values the charged axis can take. Use it according to the form of the task: mandatory for extension within a coupling, usual for completing a partial system, optional for carve-out and reimplementation, labeled honestly and never forced. Differential execution (aligning the agent's version against the original observable by observable) is used only during construction, to calibrate how complete the assertions are, and never for scoring. Otherwise it would kill implementation freedoms the specification allows (tie-breaking, internal traversal order) as if they were regressions.

Determinism. The same solution runs three times bit-identically, and the regular container pipeline agrees with the local machine test case by test case. For any task that queries the engine's navigation service or a tilemap navigation layer, asynchronous iteration of the navigation map has to be turned off in all three places - the judge side, the agent-visible side, and the recording side (avoidance tasks additionally pin avoidance to a single thread). Without that, the map rebuild runs on a wall-clock worker thread, and the same command has been measured producing frame delays of two, three, and four; with it off, the window collapses to a constant one frame. This nondeterminism enters the tested behavior itself (the agent requeries a path right after changing the terrain), not merely the observation surface, and on a long scenario "three bit-identical runs" would eventually break.

Determinism is a hard threshold, but verified isolation measures are allowed: replacing a source of nondeterminism in a single file, or black-boxing a layer that takes no part in judging. A repository that cannot be brought under control within budget drops out entirely. Resource limits on the judge side are frozen with the task, are part of the determinism and reproduction contract, and should not be modified by anyone reproducing the results; the solve-side resources may be adjusted to whatever compute is available.

Anti-cheating inside the process, in three stacked layers. The delivered module is called inside the game process, so it is in a position to bypass the tested mechanism by opening a second execution channel. All three layers read runtime state only, never the source of the delivered code.

1. The execution-channel boundary. The delivered module should run only when the criterion calls it. At judging time the runtime node tree is scanned, and any node found running the delivered code means it has mounted itself into the game and can advance on its own every frame, bypassing the prescribed call cadence. In the same spirit, the tested entity may carry only the parts its own scene declares and the ones the given code mounts on it: one extra timer amounts to building a second cadence clock, and cadence is exactly what is under test.
2. The conservation audit. At the granularity of a round or a wave, run a conservation ledger (monotonicity, frozen quantities) over runtime quantities that are not the agent's responsibility.
3. The no-regression reference does double duty. Changes that reach beyond their remit collide with the stored reference numbers.

The declared boundary is the same as it is library-wide: modifying judge internals through in-process reflection is out of scope.

Tamper protection comes from overlay order. At judging time the agent's workspace is copied out, and the task's frozen judge side is overlaid on top of it last, so any change the agent made to the level definition, the criterion, or the configuration has no effect at judging time. Producer-side tasks in a real project invert the boundary: a whitelist lets the module paths the agent owns through, and everything else is still overlaid last.

## 4. Twinned sides and a faithful preview

The judge side and the agent-visible side each hold a twinned copy of the shared simulation core - the constants, the construction of the state fed to the delivered module every frame, and the orchestration of runtime dynamics. At judging time it is overwritten by the authoritative copy. The point of twinning is that the behavior the agent sees in the preview and the behavior that drives it at judging time are the same logic, through the same interface, in a different situation.

Requirements on the agent-visible side.

- Geometry is generated procedurally at runtime from the level definition, which is its single source of truth. Do not let a tilemap stand in as the authority for geometry.
- Place one example scenario that runs immediately, as the preview.
- Keep the single implementation of the visuals in one place, and put visual logic the preview does not use in the recording layer.
- The stub for the deliverable must exist, with the interface contract in its comments. Write the stub as a visibly wrong default (settling the moment it fires, hammering a bare argmax) so the very first run surfaces feedback. Do not write it as an empty implementation. A stub comment containing the words judging or scoring is leakage as well, and is swept along with the rest.
- The preview has to be behaviorally faithful: failures (hitting a wall, timing out, dying) are visible in the preview log. One trap: if the visible entity is a physics body, a black-box query may find the entity itself, so clear its collision layer.

Stubs are very easy to forget (only the judge and the body of the project get handed in, the stub is missing, and the preview errors the moment it loads). Before handing the task in, confirm the stub is present and the preview is error-free, and confirm that build caches and resource identifier files have not slipped into the agent-visible project, since they collide with the judge-side twin during the overlay.

## 5. Landing the scenario set

- Scenario structures are designed by hand, and seeds perturb numeric values only within safe bands. Every task has exactly one example scenario, and a missing or duplicated one fails immediately at load: a structural invariant replaces a configuration knob that could drift.
- Every scored scenario declares the axis and rung it arms, which is mandatory for every scored scenario of a combo task (an atom task's judge does not consume the axis declaration and is exempt). The scenario name is only a display label; machine identity lives in the axis and rung declaration, and stacking several rungs on one axis has to report the construction honestly.
- The random stream. The example scenario uses a bare seed and has to be bit-identical with the agent-visible twin, staying plain. A scored scenario mixes the scenario name into the seed to prevent correlation across scenarios. Because the scenario name enters the random stream, a scenario is never renamed once it is in the library: renaming redraws the runtime and voids all calibration and archived readings for that cell. This is exactly why machine identity sits in the axis declaration and the name is only a label.
- An unknown or missing scenario name always fails immediately, never falls back to a guess: silently running the wrong scenario because of a typo is more dangerous than not running at all.
- The attribution contract: on a single-axis scenario the charged axis must equal the armed axis; on a coupled scenario this relaxes to the charged axis belonging to the armed set. A failure charged outside the armed set is a construction error.
- Defenses against unsolvable instances: structural legality verified by hand once; numeric bands chosen in ranges that never touch connectivity or solvability; calibration curation as the backstop (the proper solution passes everywhere with margin, the naive solution breaks on the intended axis, and unsuitable seeds do not enter the table). Nothing is random at evaluation time.

## 6. The four calibration artifacts and the calibration loop

- The proper solution uses the engine's abstractions and sequences them correctly. It must pass on every scenario at every seed, with margin, and the calibration record has to carry the margin figures. Passing narrowly on the tolerance line is forbidden: sitting on the line means an unrelated parameter shift would kill a correct solution. For carve-out and reimplementation, the proper solution is taken directly from the original upstream implementation, and the fact that it passes is itself direct evidence that the criterion does not kill correct solutions.
- The naive solution is the strongest classic-bug-pattern implementation a red-team reading would write (happy path only, countdowns running in parallel, derived quantities written straight back into base quantities, ordering once at the top of a round). It has to fail on every scored scenario, with the failure signature landing on the intended axis. Read the signature of each scored scenario separately: a mixed signature means the trap did not bite the design intent, so go back and change the scenario assignment or adjust the numeric band.
- The family of single-capability mutants removes exactly one named capability each - deleting one steering force, caching a value that should be queried every frame, collapsing a window check to its endpoints. The criterion must catch each of them; where one genuinely cannot be caught because the specification allows that implementation freedom, record the reason case by case under Section 7.
- The behavior-preserving control is a refactoring with no observable difference in the running game, or a second legal way of implementing the same mechanism. It must pass. The purpose of this reading is to keep the assertions from treating one implementation style as the correct answer, and it acts as a standing regression: the moment anyone adds an over-specific assertion back, the control fails.

The calibration loop. A proper solution that cannot pass means the criterion is too strict, so relax it. A naive solution that passes means it is not a hard gate, or the gradient is too shallow, so harden it. Avoid binary gates, and try to make a mediocre implementation separable too. What cannot be repaired is discarded.

The calibration order. Start by running a single test case locally for quick diagnosis, then sweep the full set (pass rates by scenario and by seed), and finally rerun end to end through the regular container pipeline to confirm it agrees with the local run, since determinism has to hold under isolation.

## 7. Archiving the mutant family

Mandatory for a new task, for a rebuild, and for a deepening pass.

- The mutant list plus the result matrix (mutant against scenario, pass or fail, plus the charged axis), with each mutant removing exactly one named capability;
- Every uncaught entry is disposed of individually: a genuine gap gets an assertion; an implementation freedom the specification allows (tie-breaking, internal loop order) gets its reason recorded. Nothing is left blank.
- The artifacts and the report are archived with the task, and the task configuration keeps a one-line summary.
- Whether the criterion and the calibration readings are in sync is checked against version history (the relative order of the judge and the calibration record).

The random stub is the limiting case of the mutant family, and its conclusion also serves the mechanism-ownership recheck (replace the agent-side code with a well-formed random stub and see whether the failure signature is anchored to the mechanism contract or to play outcomes).

## 8. The recording layer used for presentation

Every task also carries a recording layer that takes no part in judging, used for library presentation and documentation material. It reuses the criterion's entire simulation (never copy the simulation loop on the recording side) and only hands the settled state of each frame to the visual layer, at zero cost on the judging path. What is required is that the clip is produced and the picture is readable.

Two engineering notes. For a task whose simulation loop contains no physics wait, recording has to yield once per frame, or the whole simulation is squeezed into a single engine frame and the clip is unusable; that yield has to be gated behind recording mode so the judging path stays fully synchronous and bit-identical. For a task that queries navigation or avoidance, the determinism settings on the recording side have to match the judging side, or the behavior in the clip diverges from the behavior the criterion judged.

Recording for a real-project task keeps the original repository's visuals where possible, but the rendering layer is an audit surface for leakage as well: rendering code in the original repository often inlines mechanism logic (health-bar formulas, damage prediction lines, the conditions under which a status icon switches). The wording audit has to cover whatever rendering layer is kept, and visual code that writes the correct behavior into the picture logic is moved into the recording layer or removed, as before.

## 9. Self-check before handing in

Work through the whole acceptance standard once (the formal review is carried out by the independent review agent): the three qualification tests and the mechanism-ownership recheck; black box and the information boundary; independent recomputation of the key quantities; red-team probes and the attribution contract; determinism and agreement with the container; difficulty discipline; the mutant family archive. A real-project task adds the discoverability audit and the specification-ambiguity review.

## 10. Report template

The task name; how the three-part structure landed (judge side, agent-visible side, calibration artifacts); the calibration figures (how many test cases the proper and naive solutions passed out of how many, on the example scenario and on the scored scenarios, plus the proper solution's margin); whether discriminative power was achieved (the proper solution passes everywhere and the naive solution breaks everywhere); the mechanism-ownership finding (the three qualification tests plus the random-stub signature); the catch rate of the mutant family (how many were caught, which were not, and how each was disposed of); the traps you hit and the gaps still open. Report conclusions and the key readings. Do not paste long stretches of file content.
\end{promptbox}

\subsubsection{Review agent}
\label{app:prompts:review}
\begin{promptbox}{Review Agent}
# Review agent - instructions

## 0. Role boundary and the governing principle

This file is your only input. You do not read the authoring manual, and you do not read the reasoning sections of the blueprint or the construction notes. You judge independently against the standard written here, which is how your perspective stays isolated from the construction process.

The governing principle: take no reading reported by the construction side on trust. Every calibration figure, every pass matrix, every leakage-audit conclusion is one you obtain yourself by rerunning through the regular container pipeline. Reported readings have disagreed with measured ones more than once in this project, so this rule admits no exceptions.

What you produce is a recommended disposition, in three grades: pass, return for hardening, or fail. Whether the task enters the library is decided by a human reviewer.

## 1. The three qualification tests and the mechanism-ownership recheck (the gate for every tier)

This is the gate that decides whether a task holds up at all, and it comes before and above every individual checkpoint.

### 1.1 The three tests (all must pass)

1. Mechanism ownership. The implementation of the tested mechanic - the mechanism mathematics, meaning the integrator, the scheduler, the ledger, the solver, the state machine - must live in the code the agent delivers, rather than the agent merely scoring preferences over actions the game already offers. The test is not whether the output happens to be a choice. It is whether producing the correct output requires mechanism state on the agent's side.

The discriminator is a random stub, and it is machine-checkable: replace the agent's code with a well-formed random stub, run the criterion, and read the failure signature. If only outcome bounds break (leaks, win or loss, casualties), the mechanism lives on the game or judge side and the agent is merely a player, so the task does not qualify. If mechanism contracts break (conservation, timing, ordering), the implementation lives in the agent's hands and it qualifies.

The shorthand: delete the agent's code, and ask whether the game gets dumber (it plays badly but nothing is missing) or crippled (something stops happening altogether).

2. Where the assertions land. Failure signatures are anchored to mechanism contracts (invariants of conservation, timing, and ordering), not to play outcomes (lives left, waves cleared, whether a target was eliminated). Play outcomes serve at most as an admission gate. Whether a task qualifies lives in the design of its assertions, not in its subject matter: for the same deliverable, assertions anchored to a contract make it an engineering task, and assertions anchored to winning make it a game-playing bot.

3. Capability naming. Without naming a game, the thing under test can be stated in the vocabulary of a game-development role ("implement flocking avoidance", "implement an input buffer" qualify; if an honest description leaves nothing but "play some game well", it is a bot task).

The three-level vocabulary these tests are judged in: engine subsystems (physics, navigation, rendering, input); genre-general gameplay mechanics that have names and transfer across games; and one specific game's content systems. The tested mechanic must belong to the first two. The third may serve only as a source of difficulty (the need to read the project), never as the channel through which score is read.

### 1.2 The mechanism-ownership recheck (against drift between blueprint and build)

Mechanism on the judge side is easy to calibrate and easy to make deterministic, so the pressures of building naturally push mechanism content from the agent's side toward the judge's. The individual checkpoints inspect leakage, fairness, attribution, calibration, and determinism, but not which side the mechanism ended up on, so drift passes every light.

The rule: after the task is built and before it is admitted, you run the random stub and the three qualification tests once more, and the finding goes into the review record for the human reviewer. If a mechanism the blueprint promised on the agent's side was moved to the judge's side during implementation, that counts as a downgrade, and the recommended disposition says so honestly - change the tier assignment, withdraw the task, or send it back.

## 2. The checkpoints

Checkpoint 0 belongs to the blueprint stage; at review you verify that its evidence is on file inside the blueprint. Checkpoints 1 through 8 are worked through for every task before admission.

- Checkpoint 0, the two blueprint-stage machine checks (a prerequisite, for every new repo-tier and combo-tier blueprint): the measured evidence for the discoverability check and the empty-axis check is on file inside the blueprint (2.0).
- Checkpoint 1, black-box review: the criterion reads only observables, with no source inspection (2.1).
- Checkpoint 2, the information-boundary audit: the wording audit, the structure audit, review of the brief and the shared instruction, the disclosure in place, and discoverability (2.2).
- Checkpoint 3, independent recomputation: the criterion's key quantities are recomputed independently, never trusting what the implementation reports (2.3).
- Checkpoint 4, red-team probes: the naive solution breaks on the intended axis everywhere; margin discipline; attribution probes for coupled scenarios (2.4).
- Checkpoint 5, determinism and container agreement: three bit-identical runs, and the local machine agrees with the container pipeline test case by test case (2.5).
- Checkpoint 6, the mechanism-ownership recheck: the random stub plus the three qualification tests (Section 1).
- Checkpoint 7, difficulty discipline: mechanism-form assertions, two recipes, and the criterion-anchor check (2.7).
- Checkpoint 8, the mutant-family archive: the result matrix plus case-by-case disposition of every uncaught entry (2.8).
- Additional tests for real-project producer tasks: the three-class disposition of the discoverability audit, and the specification-ambiguity review (2.9).

### 2.0 Checkpoint 0: the two blueprint-stage machine checks

The evidence for both checks has to be produced at the blueprint stage and may not be moved to implementation or review. Both failure modes they catch - a criterion anchor that can be translated straight out of the brief, and a difficulty axis that turns out to be empty - only become visible after the task is built, and neither can be repaired then. The cost is a whole round of scenario work thrown away. Verify inside the blueprint:

1. The criterion-anchor discoverability check. Every fact the judge's assertions depend on is put through the three-class disposition, contract family by contract family (the standard is in 2.9). For the class that is discoverable in project code, independent evidence of where the fact lands has to be given, and for each family the blueprint answers: can the agent infer what has to be handled, and if it cannot, on what basis does a scored scenario charge it?
2. The empty-axis check. Every difficulty axis has one single-defect naive implementation, run across all scenarios as a measured matrix (implementation against scenario: pass or fail, plus the charged axis). That axis has to break, with zero collateral damage on other axes. This is measured, not argued. When an axis is suspected of being empty, there is a cheap discriminator: run two frozen configurations that differ only in that constant on the same situation and compare the result fingerprints.

### 2.1 Checkpoint 1: black-box review

- The criterion observes only observables of runnable behavior (positions, health, targets, attack timing, geometric clipping). No source inspection, no language model in the scoring path. Assertions are parameterized and can be rejudged under a different seed. The scored scenarios and their seeds never enter the solve environment.
- The load-bearing rule for real-project producer tasks: the criterion drives a complete game scenario for a full run, and the assertions read only runtime observables and the event stream. Never make the return value of the delivered module the object being scored, and never call it in isolation as a unit test. Breaking this degenerates the task into property-based module testing and dismantles what distinguishes this benchmark.

### 2.2 Checkpoint 2: the information-boundary audit

Everything the evaluated agent can read (the shared solve instruction used across the library, the brief, every comment in the agent-visible project) may describe only the game's functional requirements and its fixed rules, and never the evaluation mechanism. Check each item.

1. The wording audit (run over the agent-visible project, the brief, and the shared solve instruction; any hit is returned to be rewritten in neutral terms). There are four classes of word to sweep for: words naming scoring or a scorer; words naming "other situations on the evaluation side"; words naming how we measure (judging, recording, tier, scenario, and similar internal vocabulary); architectural instructions demanding robust or general code or forbidding hardcoding. A variable name holding a seed is harmless engineering vocabulary and may stay; descriptions of evaluation strategy such as "judged across seeds" or "the seed will be swapped" have to go.
2. The structure audit. Compare the key sets of the two level definitions: the agent-visible twin must not contain a key that only the judge side consumes, since the key name itself leaks the concept (in one earlier case a boolean key meaning "the door closes" was left in). Judge-side keys should be read through a default at the point of use, so the twin does not break when a key is absent.
3. The comment surface. Visual logic the preview does not use (the on-screen response to an event specific to a scored scenario), together with its comments, has to live in the recording layer rather than the agent-visible project. Comments the agent can see may only describe the behavior of the preview itself.
4. The layer-1 disclosure is present. The brief has to state the fact, about the game itself, that the arena is generated procedurally on every run and that the preview is wired to one example. It is not followed by "so make it general", and it does not mention testing.
5. Review of the shared solve instruction. It is the task-neutral operational guide used across the library, and its completion criterion has to point back to the functional requirements in the brief. The preview is positioned only as a debugging aid. There is no false criterion of the form "iterate until the preview passes", and no architectural instruction to "make it robust or general". Check the brief itself against the same points.
6. The boundary of the deliverable is in place. Nowhere does the text say "read-only" or "do not edit". What the brief calls fixed is the boundary of the deliverable (the deliverable is the implementation at the named paths; everything else is the game itself; local changes are free but are not part of the deliverable), and the second half - that the integration target is fixed - has to be present. Deleting "read-only" without establishing the boundary leaves the task ambiguous, since widening the doorway would also count as finishing.
7. The three-layer register (for settling disputes). Functional requirements and fixed rules, including the fact of generation on every run, belong to layer 1 and must be disclosed. How hard a situation is and whether anything changes mid-run belong to layer 2 and are not disclosed. The evaluation mechanism is layer 3 and is never mentioned. The dividing line is whether the text states something about the game itself or something about how we measure it. The same fact stated as a requirement is legitimate; translated into how to write the code it is forbidden.
8. Rule discoverability (in force for every tier). Every rule required for correct behavior has to be stated in the brief or be discoverable in the agent-visible project. Any scoring basis that can be found in neither is returned. The item-by-item procedure for real-project tasks is in 2.9.

### 2.3 Checkpoint 3: independent recomputation

You recompute the criterion's key quantities (geometry, timing, conservation) yourself, without trusting the numbers the implementation under test reports about itself.

### 2.4 Checkpoint 4: red-team probes and the attribution contract

- The naive solution has to fail on every scored scenario, with the failure signature landing on the intended axis. Read the signature of each scored scenario separately: a mixed signature means the trap did not bite the design intent, so it goes back for a different scenario assignment or an adjusted numeric band.
- Margin discipline. The proper solution has to pass everywhere with margin, and the calibration record has to carry the margin figures. Passing narrowly on the tolerance line is forbidden, because sitting on the line means an unrelated parameter shift would kill a correct solution.
- Three fairness checks (otherwise this is a harness defect rather than a capability test). The state interface is identical on the example scenario and the scored scenarios, and every information channel the scored scenarios rely on is offered on the example scenario too. The proper solution passes without being told, relying only on reading current runtime state. The scored scenarios only add harder situations under the same functional requirement, never a new functional requirement. Every violation the criterion punishes traces to a rule that was disclosed or is discoverable in the agent-visible project.
- Two probes for coupled scenarios. A tailored probe that is correct on axis A and naive on axis B has to be charged to B, and vice versa. A coupled scenario holds only when each axis can be attributed independently.
- The attribution contract. On a single-axis scenario the charged axis must equal the armed axis; on a coupled scenario this relaxes to the charged axis belonging to the armed set. The failure has to be charged inside the armed set, and charging outside it is a construction error.
- All-axes-armed scenarios, where present. Review is the regular checkpoints plus one probe per axis (naive on that axis, proper on the others, and it has to break on its own axis). With every trap open at once, the margin calibration of the proper solution is the largest risk, physical interference between traps has to be checked pair by pair, and the guaranteed bite of each axis has to be independent of the values drawn.

### 2.5 Checkpoint 5: determinism and agreement

- The same solution runs three times bit-identically (locally), and the local machine agrees with the regular container pipeline test case by test case.
- For a task that queries the engine's navigation service or a tilemap navigation layer, asynchronous iteration of the navigation map has to be turned off in all three places: the judge side, the agent-visible side, and the recording side (avoidance tasks additionally pin avoidance to a single thread). Without that, the map rebuild runs on a wall-clock worker thread, the same command produces several different frame delays, and "three bit-identical runs" will break on a long scenario.
- The presentation recording (which takes no part in judging) has to be produced and has to be readable.

### 2.6 Checkpoint 6: the mechanism-ownership recheck

Carried out per 1.2: the random stub plus the three qualification tests, with the finding written into the review record for the human reviewer.

### 2.7 Checkpoint 7: difficulty discipline

1. The form of the assertions has to be mechanism-form. Asserting a unique correct value or a unique legal ordering (an equality, a temporal invariant, conservation, an exact rollback) is mechanism-form and qualifies. Asserting that some bound holds (no more than N leaks, survival, coverage above a threshold) is threshold-form, may serve only as an admission gate or a saturation gradient, and may never be a hard-gate assertion used for scoring.
2. Two recipes. Difficulty draws on at least two heterogeneous recipes: engine time base and scheduling semantics; no fixed doctrine (the same policy produces opposite correct behavior in two coupled situations, with no way to know which one you are in); necessary but not sufficient in two layers (choosing the right system, then driving it correctly); undisclosed but legal call patterns (same-frame concurrency, re-entry, extreme values, a stretched time base). A single recipe collapses the gradient back into a binary gate. Whether two recipes genuinely hold is machine-checked at checkpoint 8 by the mutant family: if a task claims to press the time-base axis but a degenerate implementation that simply counts frames passes, that axis is fictional.
3. The criterion-anchor check. The correct answer the assertions anchor to must not be a one-step consequence of a rule the brief already discloses. The anchor has to sit on a dynamic situation such as a temporal snapshot or state-inheritance semantics over runtime observables, since the brief can disclose a rule family but cannot spell out the correct output for every case. A task whose anchor can be translated directly is expected to saturate: re-anchor it, admit it honestly as coverage (carrying no discriminative weight, with a difficulty alert in the calibration record), or do not build it. Tightening the wording cannot fix a misplaced anchor.
4. Depending on the engine is not the same as being difficult. Expectations are never raised on the grounds that "it uses the engine, so it must be harder".

### 2.8 Checkpoint 8: the mutant-family archive

Mandatory for a new task, for a rebuild, and for a deepening pass.

- The mutant list plus the result matrix (mutant against scenario, pass or fail, plus the charged axis), with each mutant removing exactly one named capability;
- Every uncaught entry disposed of individually: a genuine gap gets an assertion; an implementation freedom the specification allows (tie-breaking, internal loop order) gets its reason recorded. Nothing is left blank.
- The artifacts and the report archived with the task, with a one-line summary in the task configuration;
- Whether the criterion and the calibration readings are in sync, checked against version history (the relative order of the judge and the calibration record).

The random stub is the limiting case of the mutant family, and its conclusion also serves checkpoint 6. The behavior-preserving control (a refactoring with no observable difference in the running game, or a second legal way of implementing the same mechanism) has to pass: it is a standing regression, and the moment anyone adds an over-specific assertion back, the control fails.

### 2.9 Additional tests for real-project producer tasks

1. The criterion-information discoverability audit (mandatory, machine-checkable in character). List every fact the criterion's assertions depend on (constants, rules, temporal semantics), find independent evidence for it in the repository code the agent can see (skeleton and stub comments do not count, since they belong to the brief), and dispose of each under one of three classes.
- Discoverable in project code: never written in the brief, with the evidence recorded in the construction notes.
- Currently knowable only from the brief: move it back into the project where possible (keep a neighboring upstream implementation, a data table, a constant); if it cannot be moved back, it is demoted to the third class.
- Present nowhere (constants and rules that vanished along with the carve-out): it has to go into the brief as the simplest possible factual sentence, as a fallback.

No scoring basis that only the task author knows may be built on. Fix the specification or change the assertion; this line admits no exceptions. For variants built as a minimal purpose-built game, "discoverable in project code" reads as "discoverable inside the agent-visible project".
2. The specification-ambiguity review (mandatory, and it sets no implementation gate). Dispatch an independent specification-ambiguity review agent (read-only, changing nothing and judging nothing) to pick out where the specification admits more than one reading, listing them only. The list goes to the task author to settle: a genuine defect means fixing the specification, while what belongs to the need to read the project is kept and recorded. What is reviewed is the interface facts and the fallback factual sentences. "You have to read the repository to know that contract detail" is not an ambiguity; it is the design.
3. The no-regression reference is labeled honestly according to the form of the task (mandatory for extension within a coupling, usual for completing a partial system, optional for carve-out and reimplementation), never forced.

## 3. The independent rerun procedure

Order of execution.

1. Rerun every calibration artifact against every scenario and every seed through the regular container pipeline (the proper solution, the naive solution, the mutant family, the behavior-preserving control), and compare test case by test case against the construction side's calibration record. Any disagreement is returned.
2. Run the wording audit and the structure audit yourself, without reusing the construction side's scan.
3. Verify that the stub for the deliverable exists and the preview is error-free (a missing stub is a frequent omission), and that build caches and resource identifier files have not slipped into the agent-visible project.
4. Run the random stub and the targeted probes of checkpoint 4.
5. Record a finding per checkpoint and produce the review report.

## 4. Review report template

The task name; the finding for each of checkpoints 0 through 8 and the additional tests (pass, or the grounds for returning it); the calibration figures (how many test cases the proper and naive solutions passed out of how many on each scenario, plus the proper solution's margin); the mechanism-ownership finding (the three qualification tests plus the random-stub signature); the catch rate of the mutant family (how many were caught, how many were not, and how each was disposed of); the list of problems found; and the recommended disposition (pass, return for hardening, or fail). Report conclusions and the key readings. Do not paste long stretches of file content.
\end{promptbox}

\subsection{Human-annotator standards}
\label{app:prompts:human}

The two human-curated stages bracket the pipeline. Screening decides which candidate mechanics enter construction, and human review decides which finished tasks enter the library. The two documents below are what the annotators judge against. Each states the checkpoints in order, what fails a candidate at every one of them, and where a failed candidate goes.

\subsubsection{Screening}
\label{app:prompts:screening}
\begin{promptbox}{Screening}
# Screening - decision standard

The input to screening is the candidate specification sourcing produces: a description of the mechanic and an initial tier assignment, plus the license routing for real-project candidates. The output is one of three verdicts - proceed to the feasibility probe, return to the candidate pool, or reject. What is judged is whether the candidate mechanic can become a task at all, not how well it turns out once built.

This stage runs no measurements. It judges mechanisms only, and whatever cannot be settled here is left to a later stage.

Rejection records carry the mechanism, not counts. A rejected candidate stays in the pool with its cause of death and the condition under which it may return; "the trigger condition is not met yet" - an engine capability that has not been measured - is labeled separately and is not recorded as a rejection.

When in doubt, reject. Checkpoints 1 through 4 are rejection gates, and any one of them that cannot be settled with confidence returns or suspends the candidate, with the reason written down. Checkpoints 5 and 6 are judgments of degree, where stating the basis of the judgment is enough.

## 1. Checkpoint 1: can it carry a deterministic criterion (judge this first)

What to judge: whether the correctness of this mechanic can be pinned to a unique correct value, a single legal ordering of events, or a quantity the running game must conserve. One of the three has to hold.

How to judge: write one sentence stating what the hard-gate assertion looks like - "what settles at the moment of firing must be the target locked at the moment of firing", "this pool's ledger must balance entry by entry", "the phase order of these three stages is invariant under any time base". A candidate for which that sentence cannot be written does not move on.

The line between the two assertion forms:

- Mechanism-form (qualifies): asserting an equality, a temporal invariant, a conserved quantity, an exact rollback.
- Threshold-form (does not qualify as a hard gate): asserting that some bound holds - no more than N leaks, survival, coverage above a threshold, a budget not exceeded. Threshold-form may serve only as an admission gate or a saturation gradient and carries no scoring weight.

Candidates whose correctness rests on visual or experiential quality are rejected outright - how good it looks, how the controls feel, whether the pacing is comfortable. These do not even enter the pool for later reference.

## 2. Checkpoint 2: mechanism ownership and capability naming

Three qualification tests, all of which have to pass.

1. Mechanism ownership: the body of the mechanic under test - the mechanism mathematics, meaning the integrator, the scheduler, the ledger, the solver, the state machine - has to sit in the code the agent delivers, rather than the agent merely scoring preferences over actions the game already provides.
2. Where the assertions land: the failure signature has to be charged to a mechanism contract (conservation, timing, temporal invariants), not to the outcome of play (health remaining, waves cleared, whether the target died). The outcome of play may serve as an admission gate at most.
3. Capability naming: without naming a game, the task can be stated in the vocabulary of a game-development role. "Implement crowd avoidance" and "implement an input buffer" qualify; if an honest description leaves nothing but "play this game well", it is a bot task.

How to judge: this stage runs the thought experiment and the probe stage runs the measured version. Replace the code on the agent's side with a format-legal random stub and ask where the criterion breaks - if only the outcome bounds break, it does not qualify; if mechanism contracts break, it does. Shorthand: delete the agent's code, and ask whether the game gets dumber (it plays badly but nothing is missing) or crippled (something no longer happens at all). Only the latter can become a task.

The three-layer system vocabulary, used uniformly when deciding: engine subsystems (physics, navigation, rendering, input); genre-general gameplay mechanics, which are named and transferable across games, such as steering and avoidance, input buffering, production queues, line of sight, turn order, and status and modifier settlement; and the content systems of one specific game, meaning its data tables and rule instances. The mechanic under test has to belong to the first two. The third may serve as a source of difficulty, the reading wall, and never as the channel through which score is read.

Deliverable shape: the mechanic under test has to be implemented by the code the agent delivers and called by the game. A controller or bot that drives the engine to play the game is not built as a new task. What decides this is the narrative plus where the assertions land, not the function signature - a caller-shaped deliverable such as "receive state once per tick" is still a legitimate mechanism task, as long as the assertions are charged to mechanism contracts and the capability can be named in role vocabulary.

## 3. Checkpoint 3: mechanism thickness

What to judge: whether the mechanic is thick enough to be worth a task. Cross-frame state machines, temporal scheduling, and multi-quantity conservation networks come first. A thin arithmetic surface (a few clamps and sums) is accepted only where it serves a structural purpose, and the decision record then carries a difficulty alert.

Difficulty has only three positive sources: mechanism thickness, the awkwardness of legal call patterns, and the reading wall of a real project. A candidate with none of the three is not built. Mechanism surface is never trimmed on the grounds that it would be easier to implement or easier to calibrate. Engine dependence is not difficulty, and expectations are not raised on the grounds that "it uses the engine, so it is harder".

The shape prohibition: a pure-function module whose interface contract is fully given is no longer built as a new task. What this stage can judge is whether the mechanic amounts to "given the input, produce the unique output", and whether its rules would have to be written into the brief in full for the task to be fair. If both hold, the prohibition applies. A callee-shaped deliverable stays legitimate provided the contract is buried in a real project as a reading wall, or the mechanism thickness itself constitutes the implementation wall - interacting couplings and several constraints held at once, rather than table lookup and transcription. The formal check is the criterion-anchor rule at the blueprint stage.

## 4. Checkpoint 4: complete runs

What to judge: whether the game hosting this mechanic can play a run through to the end. The criterion drives a complete game scenario for a whole run, so the candidate needs a game that really starts and really finishes, whether a real repository or a purpose-built minimal game. An isolated function, or a fragment that cannot complete a run, is not built.

Real-project candidates get two further checks: the repository loads and runs in headless mode, and the license routing is already in the specification. The routing is a product of sourcing, which this stage confirms is present rather than deciding again.

## 5. Checkpoint 5: the tier assignment

Sourcing supplies the initial tier and this stage rechecks it. The three tiers:

- Atom: the minimal task around a single mechanic, the anchor for capability calibration.
- Combo: several mechanics coupled together in a purpose-built minimal game, with no reading wall from a real project. The mechanics may come from anywhere, ported from an atom task or original.
- Repo: the scope of integration into a real project. The reading wall of a real repository is a defining property, and the source of the behavioral contract is the repository itself.

What to check: a real-project candidate whose behavioral contract is to be written into the brief clause by clause is a combo task wearing a real skin, and this stage either changes its assignment or returns it. Where a combo task and a repo task are built as a pair, the reading wall and the volume of code are the only variable allowed to differ.

## 6. Checkpoint 6: a cheap library-level duplicate check

What to judge: whether an equivalent already exists in the library. Final adjudication judges this again on the finished task. Duplication is read along two dimensions:

- Capability: two tasks count as duplicates only when they test the same capability at the same rung.
- Axis lineage: the difference in readings for one axis between its atom-level isolation and a combo situation is an attribution instrument and does not count as duplication.

Genre coverage is not a reason to build. The library is laid out by functional atom rather than by game genre, so a genre with no task is not a gap, whereas a capability with no task is.

## 7. Verdicts and exits

- All six checkpoints pass: proceed to the feasibility probe, with the candidate specification as its input.
- Any of checkpoints 1 through 4 fails: reject, and record the mechanism in the pool.
- The tier assignment is wrong: change it, then judge checkpoints 3 and 6 again.
- An equivalent is already in the library: return to the pool, noting which task it duplicates. Where the two differ only in the reading wall, consider building them as a combo and repo pair instead of rejecting.
- The trigger condition is not met yet: suspend, write the condition down, and do not record it as a rejection.

## 8. Report template

The candidate name and its initial tier; a verdict per checkpoint (pass or fail, with a one-line reason); the one sentence written for checkpoint 1; the conclusions of the three qualification tests and the answer to the random-stub thought experiment; the thickness assessment, with a difficulty alert where one is needed; the conclusion of the tier recheck; the result of the library-level duplicate comparison; and the verdict - proceed to the probe, reject, suspend, or return for reassignment.
\end{promptbox}

\subsubsection{Human review}
\label{app:prompts:human_review}
\begin{promptbox}{Human Review}
# Human review - decision standard

The input to final adjudication is the report of the independent review agent, including its three-way recommendation, together with the task itself. The output is one of three verdicts: admit, return for hardening, or retire. The decision standard is the one screening works from.

The division of labor with the review agent: it judges whether the task itself holds up - whether the criterion is black-box, whether the calibration readings survive a rerun, whether the information boundary leaks, whether determinism holds. Final adjudication judges whether the task belongs in this library - whether it tests what it was meant to test, what it adds to the library, and whether it is hard enough. Both have to pass before a task is admitted.

Reading the review report: its calibration readings come from an independent rerun through the regular container pipeline and are taken as given, but check that the report carries a comparison test case by test case rather than a summarizing sentence such as "agrees with the calibration record". What it gives is a recommendation and not a verdict - a task it recommends admitting may still be retired on library-level grounds, and a task it recommends rejecting cannot be overturned.

Final adjudication does not rerun calibration. A missing reading returns the task; it is not produced here.

## 1. Checkpoint 1: does it target the intended mechanism

What to judge: whether the criterion, once built, really judges the mechanic the blueprint committed to. The primary evidence is the mechanism-ownership recheck in the review report, meaning the three qualification tests plus the random-stub signature. Check each of:

- Whether the random stub's failure signature lands on mechanism contracts or on outcome bounds. Landing on outcome bounds means the mechanism is not in fact in the agent's hands.
- Whether the assertions are charged to mechanism contracts or to the outcome of play.
- Whether the capability named is still the one named originally. A capability that became a neighboring one during construction also counts as missing the target.
- The attribution contract: a single-axis scenario requires the charged axis to equal the armed axis, and a coupled scenario relaxes this to the charged axis belonging to the armed set. A failure charged outside the armed set is a construction error.

A mechanic the blueprint placed on the agent's side and construction moved to the criterion side counts as a downgrade. The exits are to change the tier assignment, send it back, or retire it.

## 2. Checkpoint 2: library-level scope and the capability added

Scope: what is tested has to be game runtime logic, meaning whether a mechanic behaves correctly in the running game, and not general software engineering such as builds, packaging, refactoring, or dependency hygiene. The subject may be any game, but what is tested cannot slide off runtime behavior.

The capability added: against the tasks already in the library, which capability not yet covered does this one add? Duplication is read along two dimensions:

- Capability: two tasks count as duplicates only when they test the same capability at the same rung.
- Axis lineage: the difference in readings for one axis between its atom-level isolation and a combo situation is an attribution instrument and does not count as duplication.

What this stage judges is the axis the task actually arms and the defect it actually catches, not the description it was proposed under.

Genre coverage is not a reason to admit. The library is laid out by functional atom rather than by game genre, so a genre with no task is not a gap, whereas a capability with no task is.

Deliverable shape: a controller or bot that drives the engine to play the game is not admitted.

## 3. Checkpoint 3: insufficient difficulty

What to judge: whether it is hard enough, and if it is not, whether it can be deepened without changing what it measures.

- It can be deepened: return for hardening. Deepening goes through the regular procedure - reassign scenarios or adjust the numeric bands, then pass calibration and review again, with the mutant-family archive mandatory.
- Deepening would change what it measures: that is not deepening but a different task. Send it through the pipeline as a new task, or retire it.
- Deepening does not work: retire it.

Before investing in deepening, check whether the criterion anchor can be moved. An anchor sitting on a dynamic situation, a temporal snapshot or state-inheritance semantics over runtime observables, responds to tightening. An anchor resting on information that has no definition left in the repository once the system is carved out, and therefore has to be written into the brief, is structurally beyond repair, and no scenario work is spent on it. Tightening the wording cannot fix a misplaced anchor.

## 4. The three exits

- Admit: all three checkpoints pass, and no checkpoint of the review agent is left unresolved.
- Return for hardening: state what has to be strengthened, which stage does it - re-anchoring at the blueprint stage, reassigning scenarios and numeric bands at implementation, adding assertions and archiving - and which checkpoints it repeats on return. "Too easy" on its own is not a disposition.
- Retire: the mechanism goes into the candidate pool with the condition under which it may return.

## 5. Settling disputed lines

The lines below are where verdicts most often diverge. Where doubt remains, take the more conservative exit: retire over return, return over admit.

The three-layer information boundary. Functional requirements and fixed rules, including the fact that the arena is generated on every run, are layer 1 and must be disclosed. How hard a situation is and whether anything changes mid-run are layer 2 and are not disclosed. The evaluation mechanism is layer 3 and is never mentioned. The dividing line is whether the text states something about the game itself or something about how it is measured. The same fact stated as a requirement is legitimate; translated into how to write the code it is forbidden.

A harness defect against a capability test. A reasonable reading scored as zero is a harness defect, not a capability test. The usual case is a delivery boundary that was never established, where the brief dropped "read only" without stating that the integration target is fixed, so widening the doorway also counts as a way of finishing. The other is a scored scenario demanding a functional requirement that does not exist in the example scenario at all. These are returned, and never closed on the grounds that the agent should have guessed.

A charged axis is not a discovery. A failure charged outside the armed set means the task was built wrong, not that a new defect was caught.

## 6. Report template

The task name; the review agent's recommendation and whether it is adopted; a verdict per checkpoint (pass or fail, with the reason); the verification opinion on the mechanism-ownership recheck, including the random-stub signature; the library-level verdict, meaning the scope conclusion, which capability it adds, and the result of the duplicate comparison; the difficulty verdict, meaning whether it is too easy, whether it can be deepened, and the deepening path or the reason for retiring; the final verdict - admit, return for hardening, or retire; and for a returned task, what has to be strengthened and which checkpoints it repeats.
\end{promptbox}

\end{document}